\documentclass[12pt]{article}
\usepackage[font={footnotesize}]{caption}
\usepackage{amssymb}
\usepackage{amsmath}
\usepackage{xcolor}

\numberwithin{equation}{section}
\usepackage{graphicx}

\usepackage{color}
\usepackage[affil-it]{authblk}
\newcommand{\mb}[1]{{\mathbf{#1}}}
\newcommand{\rref}[1]{(\ref{#1})}
\newcommand{\bs}[1]{{\boldsymbol{#1}}}

\def\fddd#1#2{\displaystyle{\frac{\delta #1}{\delta #2}}}

\newcommand{\dsl}[1]{{\displaystyle{#1}}}
\newcommand{\RR}{{{\mathbb{R}}}}

\newcommand{\Pp}{{{\mathcal{P}}}}
\newcommand{\Ss}{{{\mathcal{S}}}}
\newcommand{\Drho}{{\Delta_\rho}}
\usepackage{graphicx}
\newcommand{\HH}{{{\mathcal{H}}}}
\newtheorem{theorem}{Theorem}[section]

\newtheorem{remark}[theorem]{Remark}

\newcommand{\ou}{{\overline{u}}}

\newcommand{\omu}{{\overline{\mu}}}
\newcommand{\oet}{{\overline{\zeta}}}
\newcommand{\orho}{{\overline{\rho}}}
\newcommand{\D}{\mathrm{d}}
\newcommand{\drho}{\rho_{{}_\Delta}}

\begin{document}
\title{{Hamiltonian aspects  of  $3$-layer stratified fluids}}
\author{R. Camassa${}^1$, G. Falqui${}^{2,4}$, G. Ortenzi${}^{2,4}$, M. Pedroni${}^{3,4}$, T. T. Vu Ho${}^2$}

\affil{
{\small ${}^1$University of North Carolina at Chapel Hill, Carolina Center for Interdisciplinary}\\ {\small Applied Mathematics,
Department of Mathematics, Chapel Hill, NC 27599, USA }\\
{\small camassa@amath.unc.edu}
\medskip\\
{\small  $^2$Department of Mathematics and Applications, 
University of  Milano-Bicocca, \\ Via Roberto Cozzi 55, I-20125 Milano, 
Italy
}\\
{\small  gregorio.falqui@unimib.it, giovanni.ortenzi@unimib.it, t.vuho@campus.unimib.it}
\medskip\\
{\small $^3$Dipartimento di Ingegneria Gestionale, dell'Informazione e della Produzione, 
\\Universit\`a di Bergamo, Viale Marconi 5, I-24044 Dalmine (BG), 
Italy}\\
{\small marco.pedroni@unibg.it}
\medskip\\
{\small  $^4$ INFN, Sezione di Milano-Bicocca, Piazza della Scienza 3, 20126 Milano, Italy}
}

\maketitle
\abstract{The theory of $3$-layer density stratified ideal fluids is examined with a view towards its generalization to the $n$-layer case. The focus is on structural properties, especially for the case of a rigid upper lid constraint. We show that the long-wave dispersionless limit is a system of quasi-linear equations that do not admit Riemann invariants. 
We equip the layer-averaged one-dimensional model with a natural Hamiltonian structure, obtained with a suitable reduction process from the continuous density stratification structure of the full two-dimensional equations proposed by Benjamin. For a a laterally unbounded fluid between horizontal rigid boundaries, the paradox about the non-conservation of horizontal total momentum is revisited, and it is shown that the pressure imbalances causing it can be intensified by  three-layer setups with respect to their two-layer counterparts. The generator of the $x$-translational symmetry in the $n$-layer setup is also identified by the appropriate Hamiltonian formalism. 
The Boussinesq limit and a family of special solutions recently introduced by de Melo Vir\'\i ssimo and Milewski are also discussed.
{\color{black}
\section{Introduction}
Density stratification is an important aspect of fluid dynamics, being inherent to a 
variety of phenomena concerning both the ocean and the atmosphere. In particular, displacement of fluid parcels from their neutral buoyancy position within a stratified flow can result in internal wave motion, whose evolution is of fundamental importance to energy propagation and distribution in both oceanic and atmospheric settings.
Simplified one-dimensional models (in particular, their quasi-linear limit) have been introduced to isolate key elements in the dynamics of these phenomena, and have been the subject of  a number of investigations, e.g., from \cite{OVS79} to the more recent \cite{Du16, Eletal17,VirMil19}, through 
\cite{CaCh96,CaCh99, LT, Chumaetal08, Chumaetal09}. 

Such simplified models are the focus of this paper. 
Two main classes of configurations have been examined in the literature: that of a free surface under constant pressure, and that of a fluid confined in a vertical channel by rigid horizontal plates. 
We first frame our discussion in more general terms by using a formalism, employed in  \cite{CaCh99} for the case $n=2$, encompassing both classes.  Within this setting, we first recall the derivation of  the dispersionless $n$-layer equations by means of the so-called hydrostatic approximation, which allows to express the pressure in terms of layer thicknesses, plus a reference (interfacial) pressure. 
The free-surface case is then briefly addressed, in particular, for the sake of concreteness, when $n=2$. For a comprehensive approach see, e.g., \cite{Choi00}, { and~\cite{GGP98,Barros06} for a discussion of a  variational approach. Extension of these results, considering
dispersive terms (as in (\ref{nlayer}) below), can be found in \cite{BGT07,PCH}.}
In this respect, our main goal is to highlight the following facts: {\it (i) the non-existence} of Riemann invariants, and {\it (ii) the existence} of a natural Hamiltonian structure}. 

To illustrate our approach, we focus on a stratified fluid composed of three 
homogeneous layers of constant density $\rho_1<\rho_2<\rho_3$, confined in a vertical channel of fixed height $h$. 
We show that the apparently paradoxical feature of 
non-conservation of horizontal momentum, already hinted at in \cite{Ben86} and thoroughly discussed in 
\cite{CCFOP12, CCFOP13} for the $2D$ Euler equations, can be detected, not unexpectedly, in 
the multiply stratified case, as also noted in \cite{VirMil19}. We notice that the added degree of freedom afforded by the $3$-layer setting makes the class of initial conditions leading to lack of 
horizontal momentum conservation much larger than its two-layer counterpart, since one can have a non-vanishing 
time-variation of the horizontal momentum {\em at the first order} in the density differences $\rho_{j+1}-\rho_{j}$ 
even with vanishing initial velocities.
%
Hamiltonian aspects of such models have also been considered, notably in \cite{BB97, CS93, CGK05}, with an emphasis on the $2$-layer case.
 We approach this problem from the setup of Hamiltonian reduction methods, and show that the Hamiltonian structure of the resulting reduced equations naturally arises via a process of reduction 
from the Hamiltonian structure  introduced by Benjamin \cite{Ben86} in the study of  
incompressible, density-stratified
Euler system in two {spatial} dimensions.
The Hamiltonian formalism allows us to derive the equations of motion as a system of conservation laws, and address  the paradox mentioned above by identifying the generator of the   $x$-translational symmetry of the problem. We present explicit initial conditions that illustrate the lack of horizontal momentum conservation, and note that, unlike the  $2$-layer case, even with zero initial velocity the horizontal momentum is not conserved. 
Next, we 
show explicitly that the reduced system does not admit Riemann invariants. 
We extend our study to the so-called Boussinesq approximation of retaining density differences for the buoyancy terms only (while disregarding density differences for the inertial terms). This limit is readily obtained 
in our canonical formalism, together with its effective Hamiltonian density and ``symmetric" solutions.}


%

\section{Sharply stratified $\boldsymbol n$-layered Euler fluids}\label{SSEF}
The incompressible Euler equations  for the velocity field $\mathbf{u}=(u,w)$ and non-constant density $\rho(x,z,t)$, in the presence of gravity 
$-g\mathbf{k}$ and with horizontal fluid domain bottom at $z=0$, are
\begin{equation}
\label{EEq}
 \frac{D \rho}{D t}=0\,, \qquad \nabla \cdot \mathbf{u} =0\,, \qquad \frac{D (\rho \mathbf{u}) }{D t} + \nabla p + \rho g \mathbf{k}=0\,, 
\end{equation}
with  boundary conditions 
$\mathbf{u}(x=\pm \infty,z,t)=\mathbf{0}$ for all $z\ge 0$ and 
$w(x,0,t)=0$ 
for all $x\in \RR$.
As usual, $D/Dt=\partial/\partial t+\mathbf{u}\cdot\nabla$ is the material derivative. A classical way to reduce the dimensionality of the model is to study the spatial averages of certain 
fields, such as velocity or density, along suitable regions, as set forth by T. Wu \cite{Wu81}. In the case of fluids stratified by gravity, where the vertical direction plays 
a  distinguished role  with respect to the other two directions, one can study the evolution of vertical means of the fields. In particular, in  \cite{Wu81}  it was established that for a density stratified fluid it holds, for each layer of thickness $\eta_i=\zeta_{i-1}-\zeta_i$,
\begin{equation}
 \int_{\zeta_{i}}^{\zeta_{i-1}} \frac{D}{Dt} f\,  \D z= \frac{\partial }{\partial t} \int_{\zeta_{i}}^{\zeta_{i-1}}  f \,  \D z+
 \frac{\partial }{\partial x} \int_{\zeta_{i}}^{\zeta_{i-1}}  u f \,  \D z\, , \qquad i=1,\dots, n,
\end{equation}
where $f(x,z,t)$ is any quantity, 
and the elevations $\zeta_\alpha$ are related with the thicknesses of the fluid's layers via 
\begin{equation}\label{etaphi}
\left\{\begin{aligned}
& \zeta_n\equiv 0\quad (\zeta_n\, \text{ is the flat bottom})\, ,\\
&\zeta_{i}=\sum_{k=0}^{n-i-1} \eta_{n-k},\quad i=1,\ldots, n-1\, .
\end{aligned} \right.
\end{equation}
Due to the intrinsic nonlinearity of the Euler equations, by applying the averaging procedure  to~(\ref{EEq}) we obtain in general a non-closed system
for the layer thicknesses  $\eta_i$ and the layer mean velocities $\ou_i$
defined as
\begin{equation}\label{lmv}
\ou_i =\dsl{\frac1{\eta_i}{\int_{\zeta_{i}}^{\zeta_{i-1}} u(x,z)\, \D z}}\, . 
\end{equation}
Indeed, additional approximations must be introduced  to close the system.  
In particular, the layer averaged equations of motion for a two-layer  incompressible Euler fluid in an infinite  channel were derived from the corresponding Euler equations in \cite{CaCh99} and extended the case of $n$ layers in \cite{Choi00}. Motions of typical wavelength $L$ were considered, under the assumptions that the ratios
\begin{equation}\label{epsi}
\epsilon=\frac{h}{L}\simeq\frac{\eta_i}{L}\,,\quad  i=1,\ldots, n\,,
\end{equation}
can be considered small. Here $h$ is the total height of the channel, while $\eta_i$  is the thickness 
of the $i$-th fluid homogeneous layer.

By using the so-called columnar ansatz (see, e.g., \cite{Wu81}), and by noticing  that equation~(\ref{epsi}), together with the incompressibility of each layer, implies that the ratio of vertical and horizontal velocities scales as $\epsilon$, in \cite{CaCh99} it was shown that the $2+1$-dimensional  Euler equations (\ref{EEq})
reduce
to the $1+1$ dimensional equations 
\begin{equation}
\label{nlayer}
 {\eta_i}_t+(\ou_i\eta_i)_x=0\,,\qquad 
 {\ou_i}_t+{\ou_i}{\ou_i}_x +(-1)^ig {\eta_i}_x + \frac{P_x}{\rho_i} + D_i =0\, ,\qquad i=1, 2\,,  
\end{equation}
where   the $D_i$ are  dispersive terms, which at $O(\epsilon^2)$ read $D_i=\frac{1}{3 \eta_i} [\eta_i^3 ({\ou_i}_{xt} +{\ou_i}{\ou_i}_{xx}-({\ou_i}_x)^2)]_x $,  
the $\ou_{i}$'s are the layer-mean velocities, $\rho_i$ is the density of the $i$-th layer, 
 and $P(x,t)$ is the interfacial pressure. 
{These results were} generalized in  \cite{Choi00, BaChMi20} to the case of $n>2$ layers, {together with the discussion of suitable conditions leading to systems of quasi-linear equations, which we summarize hereafter for the reader's convenience.


%
%
}
In fact, in this paper we shall be mainly interested in  the leading order approximation in the long-wave small parameter
series expansion, the so-called ``hydrostatic limit,'' which discards the higher order dispersive terms $D_i$ and, consistently, 
gives the following  simple expression for the pressure $p_i(x,z)$ with $z$ in the $i$-th layer, i.e., for $\zeta_{i}<z<\zeta_{i-1}$:
\begin{equation}\label{eqpress1}
\begin{split}
 p_i(x,z)&=p(x,\zeta_{i}(x))- \rho_i g (z-\zeta_{i}) \\ &= P^0- g \sum_{k=0}^{n-i-1}\rho_{n-k} \eta_{n-k} - \rho_{i} g (z-\zeta_{i})
 \, , \qquad i=1,2,\dots,n-1\, ,\\
 p_n(x,z)&=P^0-\rho_n g z\, .
\end{split}
\end{equation}
{In these relations, $P^0$ denotes an $x$ and $t$-dependent reference pressure which, without loss of generality, we can  set from now on to be the pressure at the bottom of the channel. }

In the hydrostatic approximation, the equations of motion for each layer 
contain  the averaged $x$-derivative of the pressure $\overline{{p_i}_x}$ 
(which, owing to the noncommutative property of exchanging derivative with layer-averages, is in general different from the $x$-derivative of the averaged pressure $(\overline{p_i})_x$). 
Since by  (\ref{etaphi}) $\zeta_{i}=\sum_{k=0}^{n-i-1} \eta_{n-k}$ for $i=1,\ldots, n-1$ while $\zeta_n\equiv 0$, equation \rref{eqpress1} implies that
\begin{equation}\label{px}
\begin{split}
 {{p_i}_x}=&P^0_x- g \sum_{k=0}^{n-i-1}\rho_{n-k} \eta_{{n-k}\, , x} + g \rho_{i}  (\zeta_{i})_x \\ 
 =  & P^0_x- g \sum_{k=0}^{n-i-1}(\rho_{n-k}-\rho_{i}) \eta_{{n-k}\, , x} \, , \quad i=1,\dots,n-1\, ,\\
{ p_n}_x=&P^0_x\, .
\end{split}
\end{equation}
Notice that in the hydrostatic approximation 
the $x$-derivative of the pressure 
does not depend on $z$ and therefore its layer mean is readily computed as
\begin{equation}
\begin{split}
 \overline{{p_i}_x}\equiv&\frac{1}{\eta_i}\int_{\zeta_{i}}^{\zeta_{i-1}} {p_i}_x \,  \D z \, =  {p_i}_x = 
 P^0_x- g \sum_{k=0}^{n-i-1}(\rho_{n-k}-\rho_{i}) \eta_{{n-k}\, , x}  \, , \qquad i=1,\dots,n-1\, ,\\
 \overline{ p_n}_x=&P^0_x\, .
 \end{split}
\end{equation}
We thus obtain the set of $2n$ equations
\begin{equation}
\label{mean-hydro-Euler}
\begin{split}
 &{\eta_i}_t+ (\eta_i \ou_i)_x=0 \, , \qquad i=1,\dots,n-1\,,\\
 &{\overline{u}_i}_t + \ou_i\ {\ou_i}_x+ \frac{P^0_x}{\rho_i} 
 -  g \sum_{k=0}^{n-i-1}\frac{\rho_{n-k}-\rho_i}{\rho_i} \eta_{{n-k}\, , x}=0\, ,  \qquad i=1,\dots,n-1\,, \\
&{\eta_n}_t+ (\eta_n\ou_n)_x=0\,, \\
&{\overline{u}_n}_t + \ou_n {\ou_i}_x+ \frac{P^0_x}{\rho_n}=0\, ,
\end{split}
 \end{equation}
 for the $2n+1$ dependent variables $\big(\eta_i, \ou_i, P^0\big)$. 
 
 
 \section{The free surface equations: the case $\boldsymbol{n=2}$}
\begin{figure}[t]
\centering
\includegraphics[width=13cm]{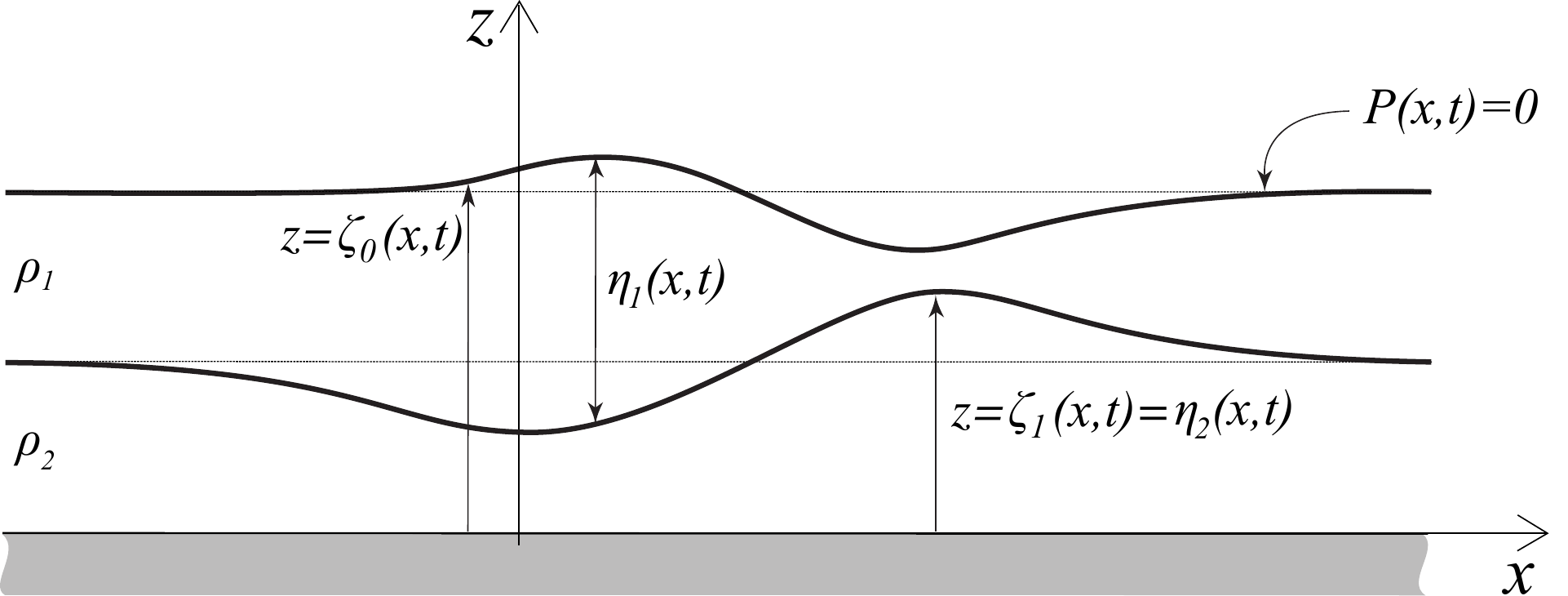}
\caption{Free-surface, two-layer fluid setup and relevant notation: $P(x,t)=0$ is the free surface (air) pressure, 
$\zeta_0$ and $\zeta_1$ are the free surface and interface locations, and $\eta_1$ and $\eta_2$ are the layer thicknesses of the fluids with densities $\rho_1$ and $\rho_2$, respectively.}
\label{Z}
\end{figure}

 A standard way to close system~(\ref{mean-hydro-Euler}) consists of considering the free surface case (see, e.g.  \cite{OVS79, Du16,Eletal17,CGK05,BLS,Choi00}). In this instance, the  {reference  pressure $P^0$} can be eliminated by observing  that the pressure at the fluid's domain upper  boundary $z=\zeta_0$ can be consistently set to vanish, as depicted in Figure \ref{Z}. Thus $P^0$ can be expressed, thanks to eq.\ (\ref{eqpress1}) for $i=1$, as
 \begin{equation}
P^0=g\,  \sum_{k=1}^n \rho_k\eta_k\,. 
\end{equation}
So, for $i=1,\ldots,n$, the 
pressures $p_{i}$ become (linear) functions of the thicknesses $\eta_{i}$  and the system closes.
An example of this procedure in the $2$-layer case is as follows (the generalisation to arbitrary $n$ being straightforward):
the pressure in the lowest layer is $p_2(x,z)=P^0-g \rho_2 z$. In the upper layer, we have 
\begin{equation}\label{pf1}
p_1(x,z)=p_2(x,\zeta_1)-g\rho_1(z-\zeta_1)=P^0-g\rho_2\eta_2-g\rho_1\eta_1, 
\end{equation}
since $\zeta_1=\eta_2$ and $\zeta_0-\zeta_1=\eta_1$.
Requiring the vanishing of the pressure at the free boundary $z=\zeta_0=\eta_1+\eta_2$ yields,
 as expected, 
$P^0=g(\rho_1\eta_1+\rho_2\eta_2)$, and so
\begin{equation}\label{press2lf}
\overline{{p_2}_x}=P^0_x=g(\rho_1{\eta_1}_x+\rho_2{\eta_2}_x),\quad \overline{{p_1}_x}=g(\rho_1{\eta_1}_x+\rho_1{\eta_2}_x)\, .
\end{equation}
Thus, the equations of the motion deduced from (\ref{mean-hydro-Euler}) are
\begin{equation}
\label{Eueq2}
\begin{array}{lcl}
 {\eta_1}_t+ (\eta_ 1\overline{u}_1)_x=0\,,  &\quad&  {\eta_2}_t+ (\eta_ 2\overline{u}_2)_x=0\,,\\ &&\\
 {\overline{u}_1}_t + \overline{u}_1\ {\overline{u}_1}_x +g\left({\eta_1}_x+{\eta_2}_x\right)=0\,, &\quad &
{\overline{u}_2}_t + \overline{u}_2\ {\overline{u}_2}_x +g \left(\dsl{\frac{\rho_1}{\rho_2}{\eta_1}_x+{\eta_2}_x}\right)=0\,.
\end{array}
 \end{equation}
These equations can be written in the standard matrix form for quasilinear systems as
\begin{equation}
\label{Meqfree2}
\left(
\begin{array}{c}
{\eta_1}_t\\
{\eta_2}_t\\
{\ou_1}_t\\
{\ou_2}_t
\end{array}
\right)+\mb{A_2}
\left(
\begin{array}{c}
{\eta_1}_x\\
{\eta_2}_x\\
{\ou_1}_x\\
{\ou_2}_x
\end{array}
\right)=0\,,
\end{equation}
where the characteristic matrix reads
\begin{equation}
\mb{A_2}=\left( \begin {array}{cccc} 
\ou_{{1}}&0&\eta_{{1}}&0\\ \noalign{\medskip}
0&\ou_{{2}}&0&\eta_{{2}}\\ \noalign{\medskip}
g&g&\ou_{{1}}&0\\ \noalign{\medskip}
\dsl{{\frac {g\rho_{{1}}}{\rho_{{2}}}}}&g&0&\ou_{{2}}\end {array} \right)\,.
\end{equation}
The question of the existence of Riemann invariants for this quasi-linear system can be easily solved by computing the so-called 
Haantjes tensor $\mathcal{H}$ of the matrix $\mb{A_2}$ \cite{H55, Bogo93, FM06}, whose vanishing ensures, in the hyperbolic case,
the existence of the Riemann invariants, and is not granted for systems with more than $2$ quasi-linear equations. 
The computation can be performed by means of standard computer algebra programs. We recover the  non-existence of Riemann invariants, conjectured in \cite{OVS79} and proved in \cite{Eletal17}, since it  
can be easily checked that the Haantjes tensor of $\mb{A_2}$ has three non-vanishing components, 
\begin{equation}
\label{HaA}
{\mathcal{H}^1}_{1\, 2}={\mathcal{H}^3}_{2\, 3}
={\frac {\eta_{{1}}{g}^{2} \left( \rho_{{1}}-\rho_{{2}} \right) }{\rho_
{{2}}}} \quad\text{and } 
{\mathcal{H}^4}_{1\, 3}= -{\frac {\eta_{{1}}{g}^{2}\rho_{{1}} \left( \rho_{{1}}-\rho_{{2}}
 \right) }{{\rho_{{2}}}^{2}}}
\, .
\end{equation}

Equations (\ref{Eueq2}) do admit a natural Hamiltonian formulation. To show this, we pass from averaged velocities $\ou_i$ to the averaged momentum coordinates $\omu_i$ defined by $\omu_i=\rho_i\ou_i$, with $i=1,2$. The evolution equations translate into
\begin{equation}
\label{pEueq2}
\begin{array}{lcl}
 {\eta_1}_t+\dsl{\frac1{\rho_1}} (\eta_ 1\omu_1)_x=0\,,  &\quad&  {\eta_2}_t+\dsl{\frac1{\rho_2}} (\eta_ 2\omu_2)_x=0\,,
 \\ \noalign{\medskip}
 {\omu_1}_t + \dsl{\frac{1}{\rho_1}} \omu_1{\omu_1}_x +g\left(\rho_1{\eta_1}_x+\rho_1{\eta_2}_x\right)=0\,, &\quad &
 {\omu_2}_t + \dsl{\frac{1}{\rho_2}} \omu_2{\omu_2}_x+g\left(\rho_1{\eta_1}_x+\rho_2{\eta_2}_x\right)=0\,.
\end{array}
 \end{equation}
 A direct inspection shows that (\ref{pEueq2}) admit the Hamiltonian formulation
\begin{equation}
\label{p2Ham}
\left(\begin{array}{c}
{\eta_1}_t\\
{\eta_2}_t\\
{\omu_1}_t\\
{\omu_2}_t
\end{array}\right)=-
\left(\begin{array}{cccc}
0&0&\partial_x&0\\
0&0&0&\partial_x\\
\partial_x&0&0&0\\
0&\partial_x&0&0
\end{array}\right)
\left(\begin{array}{c}
\delta_{\eta_{1}} H\\
\delta_{\eta_{2}}H\\
\delta_{\omu_{1}}H\\
\delta_{\omu_{2} }H
\end{array}\right)\,,
\end{equation}
where the vector $\left(\delta_{\eta_{1}} H,
\delta_{\eta_{2}}H,
\delta_{\omu_{1}}H,
\delta_{\omu_{2}} H\right)$ is the differential of the effective energy functional
\begin{equation}
\label{HfH}
H= \int_{-\infty}^{+\infty} \, \frac12\left(\frac{\eta_1}{\rho_1}\omu_1^2+\frac{\eta_2}{\rho_2}\omu_2^2
+g (\rho_1 \eta_1^2+2\rho_1\eta_1\eta_2+\rho_2\eta_2^2)\right) \D x\, .
\end{equation}
{
\begin{remark}\label{Remaref2}
{\em
The two-layer system (\ref{pEueq2}) is a non-diagonalizable Hamiltonian system of conservation laws  in Tsarev's sense~\cite{Tsa85,Tsa91}. The non-existence of Riemann invariants has deep consequences on the existence of conservation laws  and on solutions of quasi-linear systems, and as such is a topic much studied in the literature. Notable for our purposes are the results of~\cite{Tsa85,Tsa91}, showing that six quantities are guaranteed to be conserved by strong solutions: the Hamiltonian functional (\ref{HfH}),  $\int_{-\infty}^{+\infty} \eta_j\, \D x, \, \int_{-\infty}^{+\infty} \omu_j\, \D x$, $j=1,2$ (these are Casimir functionals of the Poisson brackets), and the total horizontal momentum $\Pi^{(x)}=\int_{-\infty}^{+\infty}  (\eta_1\omu_1+\eta_2\omu_2)\, \D x$, this being the  generator of the $x$-translational symmetry in the free surface case. Furthermore, in \cite{Monty} and  \cite{Barros06} the authors show that these conserved quantities are the only ones whose densities do not explicitly depend on $x$.  Also notable in this regard are the results and conjectures in~\cite{F94}, about the complete integrability of quasilinear systems which are  linearly degenerate (termed weakly nonlinear therein) but lack a complete set of Riemann invariants. 
Last but not least, as far as strong solutions of the quasi-linear system are concerned, the non-existence of Riemann invariants removes the possibility of constructing, through their level sets, a coordinate system in the hodograph space. Existence of such a system  provides a powerful tool for assuring the persistence of solutions in the hyperbolic region and for estimating times of possible shock formation.}
\end{remark}
}
 
\section{The rigid lid constraint: the case $\boldsymbol{n=3}$}\label{coperchio}
Another way of closing system~(\ref{mean-hydro-Euler}) is by  enforcing the  so-called rigid lid constraint. Its  basic physical motivation relies on the fact that surface waves effectively decouple from internal ones. This is a subtle process, since, from a physical point of view,  the interfacial hydrostatic pressure cannot be computed  {\em a priori} as in the free surface case, but rather it has to be eliminated by means of a subtraction process.

The rigid upper lid translates in the ({\em geometrical\/}) constraint
\begin{equation}
\label{geoconstr}
 \eta_1+\eta_2+\dots+\eta_n = h\, ,
\end{equation}
where $h$ is the total height of the vertical fluid ``channel". 
{ The sum of all layer-mass conservation equations ${\eta_k}_t+(\eta_k\ou_k)_x=0$ yields
 \begin{equation}
\label{neweq}
\left( \eta_1 \overline{u}_1 + \eta_2 \overline{u}_2 + \dots +\eta_n \overline{u}_n\right)_x=0 .
\end{equation}
 Taking into account the boundary conditions 
\begin{equation}
\label{bound-cond}
 \lim_{x \to \infty} \overline{u}_i =0\,, \qquad i=1,2,\dots, n\,,
\end{equation}
equation (\ref{neweq}) translates into the total flux independence of $x$ 
\begin{equation}
\label{fluxconstr}
 \eta_1 \overline{u}_1 + \eta_2 \overline{u}_2 + \dots +\eta_n \overline{u}_n=0\, , 
\end{equation}
that can be termed  the {\em dynamical\/} constraint.

The system of equations (\ref{mean-hydro-Euler}) written for the momentum densities $\eta_k\ou_k$ is
\begin{equation}\label{eqq8}
 ({\eta_i \overline{u}_i})_t + (\eta_i \overline{u}_i^2)_x+ \frac{P^0_x \eta_i }{\rho_i} 
 -  g \sum_{k=0}^{n-i-1}\frac{\rho_{n-k}-\rho_i}{\rho_i} ({\eta_{n-k}})_x \eta_i=0\,, \qquad i=1,2,\dots,n\,.
 \end{equation}
The sum of all these equations
gives the expression of the horizontal gradient of the pressure at the bottom of the channel,
\begin{equation}
\label{Pbottom}
P^0_x=-\frac{\sum_{i=1}^n(\eta_i \overline{u}_i^2)_x+g \sum_{i=1}^{n}  \sum_{k=0}^{n-i-1}\frac{\rho_{n-k}-\rho_i}{\rho_i} ({\eta_{n-k}})_x \eta_i}
  {\sum_{i=1}^{n}\eta_i /\rho_i}\,.
 \end{equation}}
 This explicit form for $P^0$ togetheer with the geometric  and dynamical constraints (\ref{geoconstr}),(\ref{fluxconstr})  guarantee the closure of~(\ref{mean-hydro-Euler}) as a system of  $2n-2$ equations for the $2n-2$ variables, say,  $(\eta_\alpha, \ou_\alpha)$ with $\alpha=2,\ldots, n$, that govern the evolution of  a sharply $n$-layer stratified fluid with upper rigid lid, i.e,, confined in a vertical channel.  
 
In analogy with the procedure we followed for the  free surface case with $2$ layers, 
rather than discussing the general case, we now concentrate our analysis to the case with {\em four} dependent variables,  that is,  to the rigid lid case with $3$ layers.

\subsection{Three-layer rigid lid case}\label{3lrl}
\begin{figure}[t]
\centering
\includegraphics[width=13cm]{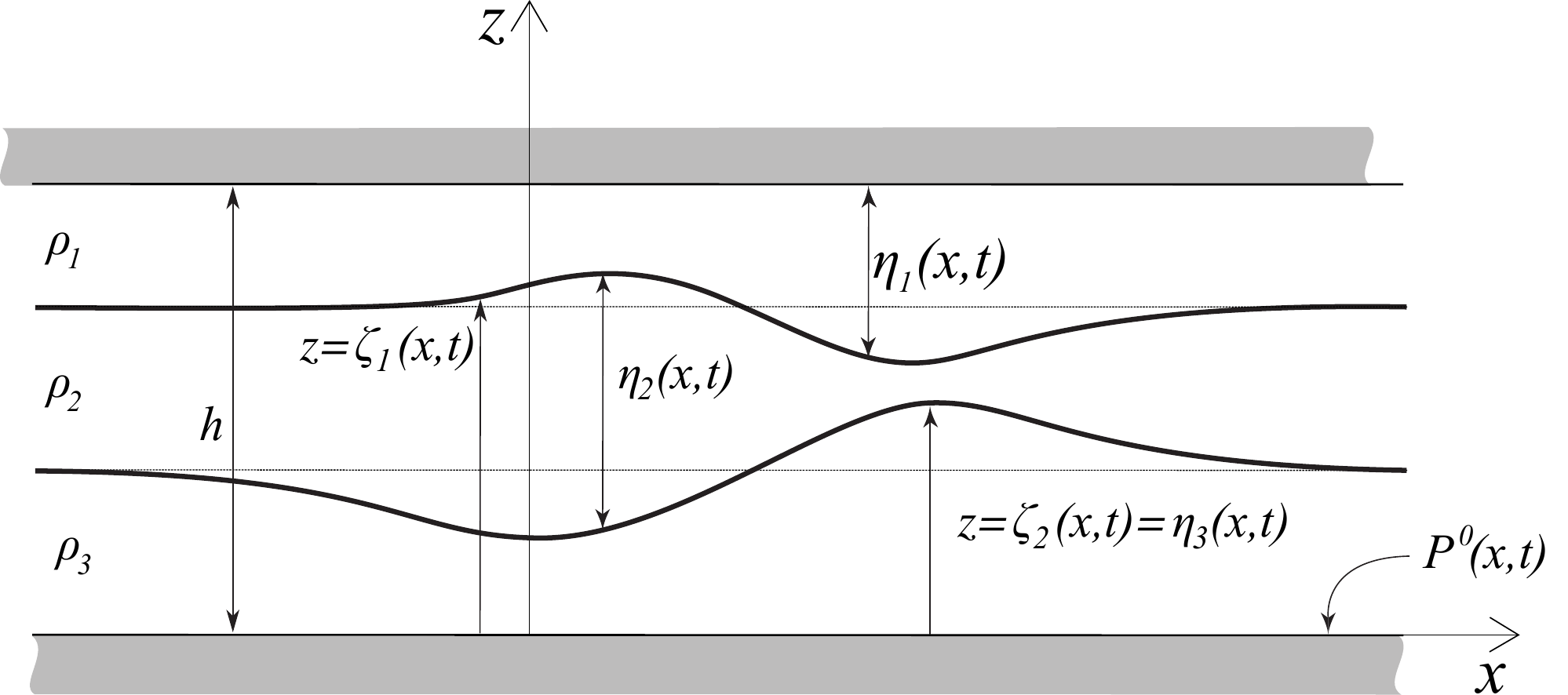}
\caption{Three-layer rigid lid 
setup and relevant notation: $P^0$ is the bottom pressure, $\zeta_i$ and $\eta_i$, $i=1,2$, are the interface locations  and layer thicknesses, respectively.}
\label{Y}
\end{figure}

Our geometrical setting is described by Figure~\ref{Y}. 
From (\ref{eqpress1}) it is clear that the explicit expression for the hydrostatic pressures in the layers is
\begin{equation}\label{3press}
\begin{split}
p_3=&P^0-g \rho_3 z \\
p_2=&P^0-g \rho_3 \zeta_2 - g \rho_2 (z-\zeta_2)=P^0-g (\rho_3-\rho_2) \zeta_2 - g \rho_2 z \\
p_1=&P^0-g \rho_3 \zeta_2 - g \rho_2 (\zeta_1-\zeta_2)- g \rho_1 (z-\zeta_1)=P^0-g (\rho_3-\rho_2) \zeta_2-g (\rho_2-\rho_1) \zeta_1 - g \rho_1 z 
\end{split}
\end{equation} 
and therefore
\begin{equation}
{p_3}_x=P^0_x\,,   \quad
{p_2}_x=P^0_x-g (\rho_3-\rho_2) {\zeta_2}_x\, ,   \quad
{p_1}_x=P^0_x-g (\rho_3-\rho_2) {\zeta_2}_x-g (\rho_2-\rho_1) {\zeta_1}_x \, .
\end{equation} 
The mass conservation equations read 
\begin{equation}\label{masscons}
{\zeta_2}_t+(\ou_3 \zeta_2)_x=0\, , \quad
{(\zeta_1-\zeta_2)}_t+(\ou_2 (\zeta_1-\zeta_2))_x=0\, , \quad
-{\zeta_1}_t+(\ou_1 (h- \zeta_1))_x=0\, , \quad
\end{equation}
which imply, together with the vanishing asymptotic conditions on the velocities, the constraint
\begin{equation}
\ou_3 \zeta_2 + \ou_2 (\zeta_1-\zeta_2)+\ou_1 (h- \zeta_1)=0\, ,
\end{equation}
as seen in \rref{fluxconstr}, since $\eta_3=\zeta_2$, $\eta_2=\zeta_1-\zeta_2$, $\eta_1=h-\zeta_1$. The Euler equations lead to
\begin{equation}\label{eqczz}
\begin{split}
&\rho_3 {\ou_3}_t+ \rho_3 \ou_3 {\ou_3}_x + P^0_x=0 \, ,\\
&\rho_2 {\ou_2}_t+ \rho_2 \ou_2 {\ou_2}_x + P^0_x-g (\rho_3-\rho_2) {\zeta_2}_x=0 \, , \\
&\rho_1 {\ou_1}_t+ \rho_1 \ou_1 {\ou_1}_x + P^0_x-g (\rho_3-\rho_2) {\zeta_2}_x-g (\rho_2-\rho_1) {\zeta_1}_x=0 \, .\\
\end{split}
\end{equation}
The mass conservation equations and the Euler equations imply the following set of evolution equations (see \rref{eqq8}):
\begin{equation}
\begin{split}
&(\eta_3 {\ou_3})_t+ (\eta_3 {\ou_3}^2 )_x + P^0_x \, \frac{\eta_3}{\rho_3}=0 \, ,\\
&(\eta_2 {\ou_2})_t+ (\eta_2 {\ou_2}^2)_x + \left(P^0_x-g (\rho_3-\rho_2) {\zeta_2}_x\right) \frac{\eta_2}{\rho_2}=0 \, , \\
&(\eta_1 {\ou_1})_t+ (\eta_1 {\ou_1}^2)_x + \left(P^0_x-g (\rho_3-\rho_2) {\zeta_2}_x-g (\rho_2-\rho_1) {\zeta_1}_x\right) \frac{\eta_1}{\rho_1}=0 \, .\\
\end{split}
\end{equation}
By taking the sum of these three evolution equations and by
using the constraint $\eta_3 \ou_3+\eta_2 \ou_2+\eta_1 \ou_1=0$, we can solve for the 
$x$-derivative of the 
pressure $P^0$ as (see~(\ref{Pbottom}))
\begin{equation}
\label{Pbottom3}
P^0_x=\frac{-(\eta_3 {\ou_3}^2+\eta_2 {\ou_2}^2+\eta_1 {\ou_1}^2 )_x+
g\left(  \frac{\rho_3-\rho_2}{\rho_2} {\eta_3}_x \eta_2  
+ \frac{\rho_3-\rho_2}{\rho_1} {\eta_3}_x \eta_1
-\frac{\rho_2-\rho_1}{\rho_1} {\eta_1}_x \eta_1
\right)}
{ \frac{\eta_3}{\rho_3}+\frac{\eta_2}{\rho_2}+\frac{\eta_1}{\rho_1}}\,,
\end{equation}
where we used again the fact that $\zeta_2=\eta_3$ and $\zeta_1=h-\eta_1$. Substituting $\eta_1=h-\eta_2-\eta_3$ yields 
\begin{equation}
\label{Pbottom4}
P^0_x=\frac{-(\eta_3 {\ou_3}^2+\eta_2 {\ou_2}^2+\eta_1 {\ou_1}^2 )_x+
g\left(  \frac{\rho_3(\rho_1-\rho_2)}{\rho_1\rho_2} {\eta_3}_x \eta_2  
+ \frac{\rho_3-\rho_1}{\rho_1} {\eta_3}_x (h-\eta_3)
+\frac{\rho_2-\rho_1}{\rho_1} {\eta_2}_x (h-\eta_2-\eta_3)
\right)}
{ \frac{1}{\rho_1}\left(h + \frac{\rho_1-\rho_2}{\rho_2}\eta_2+\frac{\rho_1-\rho_3}{\rho_3}\eta_3\right)}\,.
\end{equation}
Hence we have the following equations of  motion for $(\eta_2,\eta_3,\ou_2,\ou_3)$,
\begin{equation}
\label{Eueq2lid}
\begin{array}{lcl}
 {\eta_2}_t+ (\eta_ 2\overline{u_2})_x=0\,,  &\quad&  {\eta_3}_t+ (\eta_ 3\overline{u_3})_x=0\,,
 \\ \noalign{\medskip}
 \rho_2 {\ou_2}_t+ \rho_2 \ou_2 {\ou_2}_x + P^0_x-g (\rho_3-\rho_2) {\eta_3}_x=0\,, &\quad &
\rho_3 {\ou_3}_t+ \rho_3 \ou_3 {\ou_3}_x + P^0_x=0\,,
\end{array}
 \end{equation}
where $P^0_x$ is given by \rref{Pbottom4} and $\ou_1=-(\eta_3 \ou_3+\eta_2 \ou_2)/(h-\eta_2-\eta_3)$.
 
 \subsection{Pressure imbalances and non-conservation of horizontal  momentum }\label{Momparad}
{\color{black} Horizontal momentum conservation can be violated in the
dynamics of an ideal stratified fluid with a rigid lid constraint. This phenomenon, already suggested by Benjamin in \cite{Ben86}, was highlighted and substantiated in  \cite{CCFOP12, CCFOP13}.
The lack of horizontal momentum conservation is surprising, as the only acting body-force field is the vertical gravity, constraint forces generated through pressure against horizontal plates are necessarily vertical, and the fluid is free to move laterally. 
In the horizontal channel set-up, whenever hydrostatic conditions apply at infinity, the violation of momentum conservation is proportional (up to terms that arise from possibly different configurations at $x=\pm\infty$) to the difference $P_{\Delta}$ of the layer-averaged
pressure at the far ends of the channel. In turn, in the one-dimensional models for sharply stratified configurations such as those considered here, the layer averaged pressure imbalances are eventually those given by the quantity $P^0(x,t)$.

{
Pressure imbalances at the far-ends of the channel (see also \cite{VirMil19}), can be detected looking at equation (\ref{Pbottom3}). Indeed, its right-hand side  in {\em not} a total $x$-derivatives, and so $\int_{-\infty}^{+\infty} P^0_x\, \D x$ need not vanish even for configurations with asymptotically flat interfaces~$\zeta_i$ 
at the far-ends of the channel.
A very simple example of such a configuration with non-vanishing $P_{\Delta}$ is given in Figure~\ref{X} by splicing together arcs of parabolae for the interface locations. We choose, along with $\ou_i=0$ for  $i=1,2,3$,  
\begin{equation} \label{phiprimb}
\eta_3=\begin{cases} \frac{h}4 (1-x^2)+\frac{h}5&\mbox{if }-1<{x}<1\\ \noalign{\smallskip} \frac{h}5&\mbox{otherwise}\end{cases}\,,\qquad 
\eta_2=\begin{cases} \frac{h}3 (x^2-2\, x)+\frac{2h}5&\mbox{if } 0<{x}<2\\ \noalign{\smallskip} \frac{2h}5&\mbox{otherwise}\end{cases}\, .
\end{equation}
\begin{figure}[t]
\centering
\includegraphics[width=13cm, height=5cm]{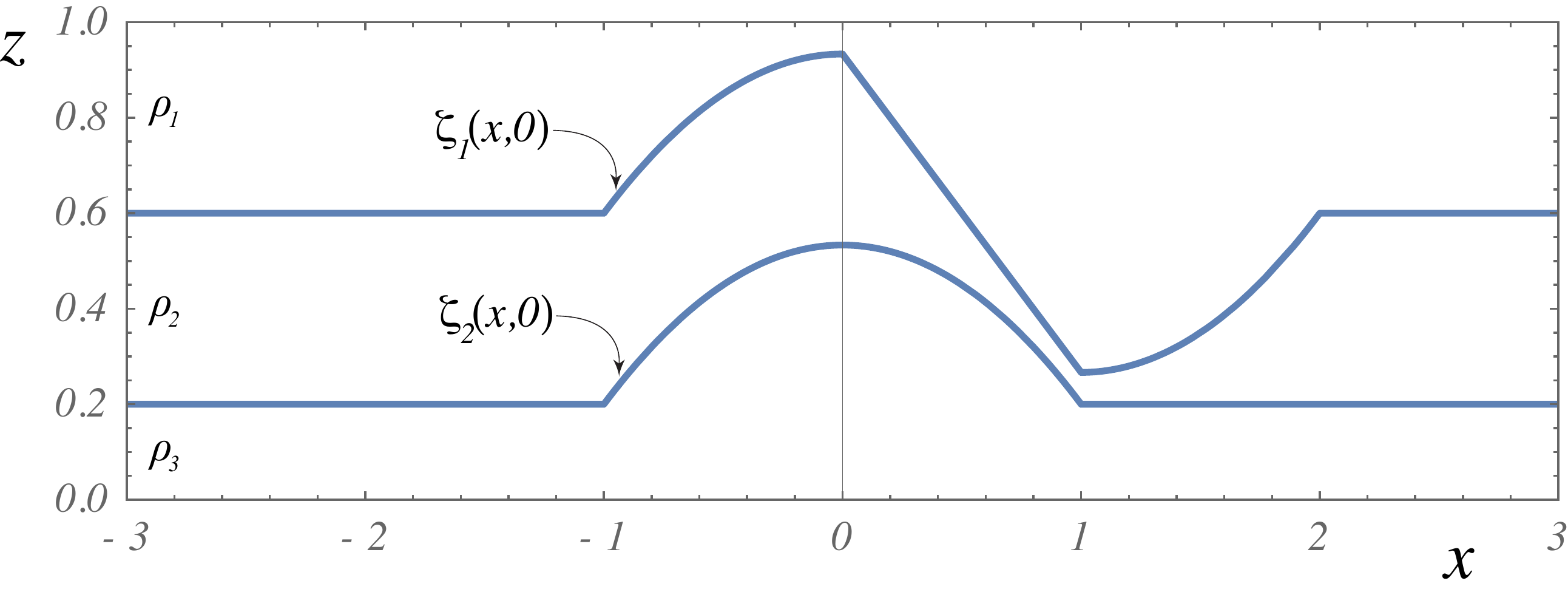}
\caption{A three-layer configuration with a non-vanishing initial pressure imbalance. Setting $\rho_1=0.5$, $\rho_2=0.75$, $\rho_3=1$, and $g=h=1$, gives $P_{\Delta}\simeq -0.0037395$.}
\label{X}
\end{figure}
A specific feature of $3$- (and, {a fortiori}, $n$-) layered fluid configurations is that this ``paradox" about non-conservation of horizontal momentum is an effect of order one in the single density differences $\rho_2-\rho_1$ and $\rho_3-\rho_2$, even with initial null velocities, as opposed to $2$-layer, rigid lid configurations, where
non-conservation of the horizontal total momentum is, with null-velocities, {\em quadratic} in the density difference $\rho_2-\rho_1$ (see \cite{CCFOP13}).
Indeed, 
equation \rref{Pbottom4} reduces to 
\begin{equation}
\label{Pbottom0}
P^0_x=\frac{
g\left(  \frac{\rho_3(\rho_1-\rho_2)}{\rho_1\rho_2} {\eta_3}_x \eta_2  
+ \frac{\rho_3-\rho_1}{\rho_1} {\eta_3}_x (h-\eta_3)
+\frac{\rho_2-\rho_1}{\rho_1} {\eta_2}_x (h-\eta_2-\eta_3)
\right)}
{ \frac{1}{\rho_1}\left(h + \frac{\rho_1-\rho_2}{\rho_2}\eta_2+\frac{\rho_1-\rho_3}{\rho_3}\eta_3\right)}\,.
\end{equation}
The numerator 
is equal, up to total derivatives, to 
$
g
(\rho_2-\rho_1) (\rho_3-\rho_2){\eta_3}_x\eta_2/(\rho_1\rho_2)$, 
thus revealing  the leading order dependence of $P_\Delta$ on the density differences
for general initial layer thicknesses $\eta_i$.
}
Finally, a quick glance at the expression of $P^0_x$ for the $n$-layer case given by equation~(\ref{Pbottom}) shows that the lack of conservation of the horizontal momentum persists for $n>3$.}
{
\begin{remark} {\em Non-dispersive, quasilinear  equations governing stratified flows are generically 
systems of mixed type~\cite{Chumaetal09, BoMi11}, i.e., exhibit hyperbolic to elliptic transitions corresponding to eigenvalues of the characteristic matrix becoming complex across some ``sonic" surfaces. The presence of elliptic domains is associated with instabilities, and, as well known,  the evolution from initial values in the elliptic domain is ill-posed. One can prove that the initial data with initial null velocities, and with initial profiles of the kind exemplified by equation~(\ref{phiprimb}), with density ratios $\rho_2/\rho_3=3/4, \rho_1/\rho_3=1/2$, lie inside the hyperbolic region, as the characteristic matrix has four real eigenvalues. By continuity, the system's evolution must remain hyperbolic for a possibly small but non-zero finite time. In fact, numerical integration of this initial value problem (not reported here) by using the reduced system derived below in \S\ref{redham}, indicates that the evolution remains hyperbolic for a relatively long time, at least till the final time $t=5$ we have tried,  with a shock forming at around time $t\simeq 2.8$ (with the nondimensional parameters as in the example of Figure~\ref{X}).}
\end{remark}}

{
\section{The 3-layer rigid lid case: the Hamiltonian structure}\label{Hamred}
In this section, we extend our study of the $3$-layer rigid-lid system considered in \S\ref{3lrl}
with the aim of endowing its equations of motion
with a Hamiltonian structure. In the free surface 2-layer case (as well as in the free surface 3-layer case, see the Appendix) the resulting equations, when written in the momentum coordinates $\omu_k=\rho_k\ou_k, k=1,2,3$, are sufficiently simple to be cast in Hamiltonian form at a glance. This is not the case in the presence of a rigid top lid, basically owing to the nonlinear nature of  the dynamical constraint (\ref{fluxconstr}). 
A natural avenue of attack to account for this is by means of a suitable Hamiltonian reduction of the Poisson 
structure for continuously stratified flows, originally introduced in \cite{Ben86},} by extending to the multiple layer case a technique introduced in~\cite{CFO17}.
\subsection{The 2D Benjamin model for heterogeneous fluids in a channel}\label{Sect-1}
Benjamin \cite{Ben86} proposed and discussed a  set-up for the Hamiltonian formulation of an incompressible stratified Euler system in $2$ spatial dimension, 
which we  hereafter summarize for the reader's convenience.
 
Benjamin's idea was to consider, as  basic variables for the evolution of a perfect inviscid and incompressible but heterogeneous fluid in 2D, 
subject to gravity, the density $\rho$ together with the ``weighted vorticity" $\Sigma$ defined by
\begin{equation}
\label{sigmadef}
\Sigma=\nabla\times (\rho\bs{u})=(\rho w)_x-(\rho u)_z. 
\end{equation}
The equations of motion for these two fields, ensuing from (\ref{EEq}), are 
\begin{equation}
\label{eqsr}
\begin{array}{l}
\rho_t+u\rho_x+w\rho_z =0\\
\noalign{\medskip}
\Sigma_t+u\Sigma_x +w\Sigma_z +\rho_x\big(gz-\frac12(u^2+w^2)\big)_z+\frac12\rho_z\big(u^2+w^2\big)_x=0\ .
\end{array}
\end{equation}
They can be written in the form
\begin{equation}
 \label{heq}
{\rho_t}=-\left[\rho,  \dsl{\fddd{H}{\Sigma}}\right] \, , \qquad 
\Sigma_t= -\left[\rho,  \dsl{\fddd{H}{\rho}}\right]-\left[\Sigma, \dsl{\fddd{H}{\Sigma}}\right] \, ,
\end{equation}
where, by definition, $[A, B] \equiv A_xB_z-A_zB_x$, and the  functional 
\begin{equation}
\label{ham-ben}
H= \dsl{\int_\mathcal{D} \rho\left(\frac12 |\bs{u} |^2+g z\right)\,{\rm d}x\,{\rm d}z }
\end{equation}
is simply given by the sum of the kinetic and potential energy, $\mathcal{D}$ being the fluid domain.
The most relevant feature of this coordinate choice is that $(\rho,\Sigma)$ are physical variables.
Their use, though confined to the 2D case with the above definitions, allows one to avoid  the introduction of Clebsch 
variables (and the corresponding subtleties associated with gauge invariance of the Clebsch potentials) 
which are often used in the Hamiltonian formulation of both 2D and the general $3D$ case (see, e.g.,  \cite{Z85}).

As shown by Benjamin, equations (\ref{heq}) are a Hamiltonian system with respect to a 
Lie-theoretic Hamiltonian structure, that is, 
they can be written as
\[
 \rho_t=\{\rho, H\}
,\qquad \Sigma_t=\{\Sigma, H\}
\]
for the Poisson bracket defined by the Hamiltonian operator
\begin{equation}\label{B-pb}
J_B=-
\left(\begin{array}{cc}
       0 & \rho_x \partial_z -\rho_z \partial_x \\ 
       \rho_x \partial_z -\rho_z \partial_x & \Sigma_x \partial_z -\Sigma_z \partial_x
      \end{array}
\right).
\end{equation}
\subsection{The reduction process}


By means of the Heaviside $\theta$  and  Dirac $\delta$ generalized functions, a three-layer fluid configuration can be introduced with 
constant densities $\rho_i$  and velocity components $u_i(x,z)$, $w_i(x,z)$, $i=1,2,3$   
(for the upper  $i=1$, middle $i=2$, and lower layer $i=3$, respectively),  with interfaces $\zeta_1$ and $\zeta_2$.
The global density and velocity variables can be written as 
\begin{equation}\label{rhouw}\begin{split}
&\rho (x,z)=\rho_{3}+(\rho_{2}-\rho_{3})\theta(z-\zeta_{2})+(\rho_{1}-\rho_{2})\theta(z-\zeta_{1}) \\
&u (x,z)=u_{3}+(u_{2}-u_{3})\theta(z-\zeta_{2})+(u_{1}-u_{2})\theta(z-\zeta_{1}) \\
&w (x,z)=w_{3}+(w_{2}-w_{3})\theta(z-\zeta_{2})+(w_{1}-w_{2})\theta(z-\zeta_{1})\, . \end{split}
\end{equation}
Thus, the density-weighted vorticity $\Sigma =(\rho w)_{x}-(\rho u)_{z}$ is 
\begin{equation}\begin{split}
    \Sigma=&\rho_3({w_3}_x-{u_3}_z)+\theta(z-\zeta_{2})\left(\rho_{2}{w_2}_x-\rho_{2}{u_2}_z+\rho_{3}{u_3}_z-\rho_{3}{w_3}_x\right)\\ \nonumber
    &+\theta(z-\zeta_{1})\left( \rho_{1}{w_1}_x-\rho_{1}{u_1}_z+\rho_{2}{u_2}_z-\rho_{2}{w_2}_x \right) \\\nonumber
    &+\delta(z-\zeta_{2})\left( (\rho_{3}w_{3}-\rho_{2}w_{2}){\zeta_2}_x+(\rho_{3}u_{3}-\rho_{2}u_{2}) \right) \\\nonumber
    &+\delta(z-\zeta_{1}) \left( (\rho_{2}w_{2}-\rho_{1}w_{1}){\zeta_1}_x+(\rho_{2}u_{2}-\rho_{1}u_{1}) \right) \\\nonumber
    =&\rho_{3}\Omega_{3}+\theta (z-\zeta_{2})(\rho_{2}\Omega_{2}-\rho_{3}\Omega_{3})+\theta (z-\zeta_{1})(\rho_{1}\Omega_{1}-\rho_{2}\Omega_{2})\\\nonumber
    &+\left( (\rho_{3}u_{3}-\rho_{2}u_{2})+(\rho_{3}w_{3}-\rho_{2}w_{2}){\zeta_2}_x\right)\delta (z-\zeta_{2}) \\\nonumber
    &+\left( (\rho_{2}u_{2}-\rho_{1}u_{1})+(\rho_{2}w_{2}-\rho_{1}w_{1}){\zeta_1}_x\right)\delta (z-\zeta_{1})\,,
\end{split}
\end{equation}
where $\Omega_{i}={w_i}_x-{u_i}_z$ for $i= 1,2,3$.
Next, we assume the motion in each layer to be irrotational, so that $\Omega_{i}=0 $ for all $ i=1,2,3$. Therefore the density weighted vorticity acquires the form
\begin{align}
\label{sigma-ham}
    \Sigma=&\left((\rho_{3}u_{3}-\rho_{2}u_{2})+(\rho_{3}w_{3}-\rho_{2}w_{2}){\zeta_2}_x \right) \delta(z-\zeta_{2}) \\ \nonumber
    &+\left((\rho_{2}u_{2}-\rho_{1}u_{1})+(\rho_{2}w_{2}-\rho_{1}w_{1})\zeta_{1x} \right) \delta(z-\zeta_{1})\,.
\end{align}
In the long-wave asymptotics described in Section \ref{SSEF}, we have, at the leading order in the long wave expansion parameter, 
$\epsilon=
h/L\ll 1$, 
$$u_{i} \sim {\ou_{i}}\,,\qquad w_{i} \sim 0\,, $$
i.e., we can neglect the vertical velocities $w_{i}$ and trade the horizontal velocities $u_{i}$ with their layer-averaged counterparts.
Thus, from \rref{sigma-ham} and recalling the first of (\ref{rhouw}), we obtain 
\begin{equation}\label{I-sigma-rho}
\begin{split}
     \rho (x,z)&=\rho_{3}+(\rho_{2}-\rho_{3})\theta(z-\zeta_{2})+(\rho_{1}-\rho_{2})\theta(z-\zeta_{1})\\
     \noalign{\medskip}
     \Sigma(x,z)&=\left(\rho_{3}{\ou_{3}}-\rho_{2}{\ou_{2}} \right)
     \delta(z-\zeta_{2}) +\left(\rho_{2}{\ou_{2}}-\rho_{1}{\ou_{1}} \right) \delta(z-\zeta_{1}) 
     \, .
     \end{split}
\end{equation}
The $x$ and $z$-derivative of the Benjamin's variables given by equations (\ref{I-sigma-rho}) are 
generalized functions supported at the surfaces   $\{z=\zeta_1\}\cup \{z=\zeta_2\}$, and are
computed as
\begin{equation}\label{rhovar}\begin{split}
    \rho_{x}=&-(\rho_{2}-\rho_{3}){\zeta_2}_x \delta (z-\zeta_{2})-(\rho_{1}-\rho_{2}){\zeta_1}_x \delta (z-\zeta_{1})  \\
    \rho_{z}=&(\rho_{2}-\rho_{3})\delta (z-\zeta_{2})+(\rho_{1}-\rho_{2})\delta (z-\zeta_{1})\, ,
\end{split}
\end{equation}
and
\begin{equation}\label{dsigma}\begin{split}
    \Sigma_{z}=&(\rho_{3}{\ou_{3}}-\rho_{2}{\ou_{2}}) \delta'(z-\zeta_{2})
    +(\rho_{2}{\ou_{2}}-\rho_{1}{\ou_{1}})\delta'(z-\zeta_{1}) \\
    \Sigma_{x}=&-(\rho_{3}{\ou_{3}}-\rho_{2}{\ou_{2}})\delta'(z-\zeta_{2}){\zeta_2}_x 
    -(\rho_{2}{\ou_{2}}-\rho_{1}{\ou_{1}})\delta'(z-\zeta_{1}){\zeta_1}_x\\ 
    &+\left(\rho_{3}{{\ou_3}_x}-\rho_{2}{{\ou_2}_x} \right)
     \delta(z-\zeta_{2}) +\left(\rho_{2}{{\ou_2}_x}-\rho_{1}{{\ou_1}_x} \right) \delta(z-\zeta_{1})\, .
\end{split}\end{equation}
To invert the map (\ref{I-sigma-rho}) we can integrate along the vertical direction $z$. Contrary to $2$-layer case described in \cite{CFO17}, now we  need to integrate on two vertical slices of the channel in order to obtain four $x$-dependent fields.
To this end, we extend the Benjamin's manifold $\mathcal{M}$ (parametrized by $\rho(x,z)$ and $\Sigma(x,z)$)
by  the space $\mathcal{F}$ of isopycnals $z=f(x)$, 
and trivially extend by $0$ the Poisson structure $J_B$ given by \rref{B-pb}. 

As shown in \cite{CCFOP14}, the operator~(\ref{B-pb})  provides a proper Poisson structure under the condition that the density $\rho$ be constant
at the physical boundaries $z=0$ and $z=h$. 
Once equipped with the isopycnal $z=f(x)$,
we can define a projection $\pi: \widetilde{\mathcal{M}}\equiv \mathcal{M}\times \mathcal{F} \to (C^\infty(\RR))^4$  by means of  
\begin{equation}\label{xitaugen}
\begin{split}
&\pi\left(\rho(x,z), \Sigma(x,z), f \right)=\left(\xi_{1},\xi_2,\tau_1,\tau_2\right)\\=&
\left ( \int_{0}^{h} (\rho (x,z)-\rho_{max} )\, \D z, \int_{0}^{f} (\rho (x,z) -\rho_{max} )\, \D z,
    \int_{0}^{h} \Sigma (x,z) \D z,\int_{0}^{f} \Sigma (x,z) \D z\right)\, .  \end{split}
   \end{equation}
 We choose to include the manifold of three-layer fluid configurations in the space $\mathcal{M}\times \mathcal{F}$ by 
 adding to (\ref{I-sigma-rho}) the
 following choice for the isopycnal $f$:
\begin{equation}\label{zetabar}
f=\dsl{{\frac{\zeta_{1}+\zeta_{2}}{2}}}\, ,
\end{equation}
hereafter denoted by $ \oet$.
Equations (\ref{I-sigma-rho}) and (\ref{zetabar})  define a submanifold $\mathcal{I}$ of $\widetilde{\mathcal{M}}$,  in one-to-one correspondence 
with the manifold
$\mathcal{S}$ of 3-layer fluid configurations considered in Section \ref{3lrl},  
%
as it is  parametrised by four functions of the horizontal coordinate $x$, and this set is equivalent to the $4$-tuple $(\eta_1,\eta_2,\ou_1,\ou_2)$.
For further reference, we explicitly remark that on $ \mathcal{I}$ 
the projection $\pi$ is defined by the 
$4$-tuple
\begin{equation}\label{xitau}
\pi\left(\rho(x,z), \Sigma(x,z), \oet \right)=\left ( \int_{0}^{h} (\rho (x,z)-\rho_3 )\, \D z, \int_{0}^{\oet} (\rho (x,z) -\rho_3 )\, \D z,
    \int_{0}^{h} \Sigma (x,z) \D z,\int_{0}^{\oet} \Sigma (x,z) \D z\right)\, \, .
\end{equation}
Further, as we shall see below, the $\dot{\bar{\zeta}}$ component of the tangent vector $(\dot{\rho},\dot{\Sigma},\dot{\bar{\zeta}})$ can be lineraly expressed in terms of $\dot{\rho}$.

To obtain a Hamiltonian structure on the manifold $\mathcal{S}$ by reducing Benjamin's parent structure~(\ref{B-pb}), we perform the following steps:
\begin{enumerate}
\item Starting from a 1-form on the 
manifold $\mathcal{S}$, represented by the 4-tuple $(\alpha_S^{(1)}, \alpha_S^{(2)},\alpha_S^{(3)},\alpha_S^{(4)})$,
we construct its pull-back  to 
$\mathcal{I}$, that is, the 1-form $\boldsymbol{\alpha}_M=(\alpha_M^{(1)},\alpha_M^{(2)}, 0)$  at $T^*\widetilde{M}\vert_\mathcal{I}$ satisfying the relation
\begin{equation}\label{eqlift}
\int_{-\infty}^{+\infty}\int_0^h( \alpha_{M}^{(1)} \dot{\rho}+\alpha_{M}^{(2)}\dot{\Sigma})\, \D x\, \D z
= \int_{-\infty}^{+\infty}
\sum_{k=1}^4\alpha_{S}^{(k)} \cdot \left(\pi_{*} (\dot{\rho},\dot{\Sigma})\right)^k \, \D x\>\>,
\end{equation}
where $\pi_*$ is the tangent map to (\ref{xitau}). 

\item
We apply Benjamin's operator~(\ref{B-pb}) to the lifted one form $\boldsymbol{\alpha}_M$ 
to get the vector field
\begin{equation}\label{Yfields} \mb{Y}\equiv
\left (\begin{array}{c} Y_{M}^{(1)}\\
Y_{M}^{(2)}\\0 \end{array}\right)= J_B \cdot \left (\begin{array}{c} 
\alpha_M^{(1)}\\\alpha_M^{(2)}\\ 0
\end{array}\right) 
\, . 
\end{equation}
\item 
Thanks to the form of Benjamin's Poisson structure, it turns out that $\mb{Y}$ 
is still supported on the
locus $\{z=\zeta_1\}\cup \{z=\zeta_2\}$, and can be easily projected with
 $\pi_*$ to obtain the vector field $(X_1,X_2,X_3,X_4)$ on $\mathcal{S}$. The latter depends linearly on $\{\alpha_S^{(i)}\}_{i=1,\ldots,4}$, and defines
 the reduced Poisson operator $\mathcal{P}$ on $\mathcal{S}$. 
\end{enumerate}
This construction essentially works as in the two-layer case considered in \cite{CFO17}, provided one  point is taken into account, namely that the integrals in the  second and fourth  component of $\pi$ have a variable upper bound. We have, 
for $(\dot{\rho},\dot{\Sigma})\in TM\big\vert_\mathcal{I}$, 
\begin{equation}\label{pistar1}
\pi_{*}\left(
\begin{array}{c} 
\dot{\rho}\\ \dot{\Sigma}\\ \dot{\overline{\zeta}} \end{array}
\right)
=
\left(
\begin{array}{l}\medskip
\int_0^h \dot{\rho}\, \D z\\ \medskip
\int_0^{\overline{\zeta}} \dot{\rho} \, \D z+\dot{\overline{\zeta}}\,(\rho(x, \overline{\zeta})-\rho_3)\\ \medskip
\int_0^h \dot{\Sigma}\, \D z\\ \medskip
\int_0^{\overline{\zeta}} \dot{\Sigma} \, \D z+\dot{\overline{\zeta}}\, \Sigma(x, \overline{\zeta}) \,.
\end{array} 
\right)\,.
\end{equation}
We remark that, since  the inequalities  $\zeta_2<\overline{\zeta}\equiv\dsl{\frac{\zeta_1+\zeta_2}{2}}<\zeta_1$  hold in the strict sense, the second term of 
this vector's fourth component vanishes. The same cannot be said for the analogous term in the vector's second component, since $\rho(x, \overline{\zeta})-\rho_3=\rho_2-\rho_3\neq 0$.
As anticipated above, on $\mathcal{I}$  the mean of the tangent vector component coming from the $\zeta$'s,
$\dot{\bar{\zeta}}=
(\dot{\zeta}_1+\dot{\zeta}_2)/2$, 
can be expressed in terms of the tangent vector component~$\dot{\rho}$. 
To this end, we can use the analogue of relations~(\ref{rhovar}), 
which generically give
\begin{equation}
\label{rhodot}
\dot{\rho}=(\rho_3-\rho_2)\dot{\zeta}_2\delta(z-\zeta_2)+(\rho_2-\rho_1)\dot{\zeta}_1\delta(z-\zeta_1)\, .
\end{equation}
Integrating this with respect to $z$ in $[0,h]$ yields
\begin{equation}
\label{A1}
\int_0^h \dot{\rho}\, \D z=(\rho_3-\rho_2)\dot{\zeta}_2+(\rho_2-\rho_1)\dot{\zeta}_1\, ,
\end{equation}
while by integrating over $[0, \overline{\zeta}]$ we obtain
\begin{equation}
\label{A2}
\int_0^{\overline{\zeta}} \dot{\rho}\, \D z=(\rho_3-\rho_2)\dot{\zeta}_2\, .
\end{equation}
Solving the linear system given by (\ref{A1}) and (\ref{A2}), gives
\begin{equation}
 (\rho_2-\rho_3)\dot{\overline{\zeta}}=\Delta_\rho\int_0^h\dot{\rho}\, \D z-(\frac12+\Delta_\rho)\int_0^{\overline{\zeta}} \dot{\rho}\, \D z\, ,
\end{equation}
where 
\begin{equation}\label{Drrhodef}
\Delta_\rho:=\dsl{
\frac{\rho_2-\rho_3}{2(\rho_2-\rho_1)}}\, .\end{equation}
Thus, formula (\ref{pistar1}) for the components of the push forward $\pi_*$ does not explicitly depend on the last component 
$\dot{\overline{\zeta}}$, and
can be written as
 \begin{equation}\label{pistar}
\pi_{*}\left(
\begin{array}{c} 
\dot{\rho}\\ \dot{\Sigma}\\ \dot{\overline{\zeta}}\end{array}
\right)
=
\left(
\begin{array}{l}\medskip
\int_0^h \dot{\rho}\, \D z\\ \medskip
\Delta_\rho\int_0^h \dot{\rho}\, \D z +(\frac12-\Delta_\rho) \int_0^{\overline{\zeta}} \dot{\rho} \, \D z\\ \medskip
\int_0^h \dot{\Sigma}\, \D z\\ \medskip
\int_0^{\overline{\zeta}} \dot{\Sigma} \, \D z
\end{array} 
\right) \,.
\end{equation}
By substituting this result in relation (\ref{eqlift}) we find
\begin{equation}\label{lift2}\begin{split}
&\int_{-\infty}^{+\infty} \int_0^h (\dot{\rho}\, \alpha^{(1)}_M+\dot{\Sigma}\, \alpha^{(2)}_M)\, \D x\, \D z\\ &=
\int_{-\infty}^{+\infty} \left(\int_0^h \dot{\rho} (\alpha^{(1)}_S+\Delta_\rho\alpha^{(2)}_S)\, \D z+\right. 
\left. \int_0^{\overline{\zeta}}\, \dot{\rho}\, (\frac12-\Delta_\rho)
  \alpha^{(2)}_S)  
\,\D z+\int_0^h \dot{\Sigma}\alpha^{(3)}_S\, \D z+\int_0^{\overline{\zeta}} \dot{\Sigma}\alpha^{(4)}_S \, \D z\, \right)\D x\, .\end{split}
\end{equation}
Hence,  the non-vanishing  components of the 1-form $\bs{\alpha}_M$ pulled back to $\mathcal{I}$ are
\begin{equation}
\label{alphalist}
\alpha^{(1)}_M=
\alpha^{(1)}_S+\Delta_\rho \alpha^{(2)}_S\theta(\oet-z) +(\frac12 -\Delta_\rho)\alpha^{(2)}_S
\,,\qquad
\alpha^{(2)}_M=
\alpha^{(3)}_S+\alpha^{(4)}_S\theta(\oet-z)\,.
\end{equation}
Applying the Poisson tensor \rref{B-pb} to this 1-form yields, after some manipulations that crucially use the fact that products of 
Dirac's $\delta$ 
supported at different locations can be consistently set to zero,  the vector field
\begin{equation}
\label{Y12} \begin{split}
Y_{M}^{(1)}=&\big((\rho_{2}-\rho_{3})\delta(z-\zeta_{2})+(\rho_{1}-\rho_{2})\delta(z-\zeta_{1})\big)\, \alpha_{S,x}^{(3)}
+(\rho_{2}-\rho_{3})\delta(z-\zeta_{2})\, \alpha_{S,x}^{(4)}\\
Y_{M}^{(2)}=&\big((\rho_{2}-\rho_{3})\delta(z-\zeta_{2})+(\rho_{1}-\rho_{2})\delta(z-\zeta_{1})\big)\alpha_{S,x}^{(1)}\\ &
+\big(\frac12(\rho_2-\rho_3)\delta(z-\zeta_2)+(\rho_1-\rho_2)\Drho\delta(z-\zeta_1)\big)\alpha_{S,x}^{(2)}
%
\\ 
    &+\big((\tau_1-\tau_2)\delta'(z-\zeta_{1}){\zeta_1}_x +
     \tau_2\delta'(z-\zeta_{2}){\zeta_2}_x \big)\alpha_{S,x}^{(3)} \,.
\end{split}
\end{equation}
The push-forward under the map $\pi_*$ of this vector field gives the following four components:
\begin{equation}\label{pushfwd}
\begin{split}
\int_0^h Y_{M}^{(1)}\, \D z&=(\rho_1-\rho_3)\alpha_{S,x}^{(3)}+(\rho_{2}-\rho_{3})\alpha_{S,x}^{(4)}\\ 
\Delta_\rho\int_0^h Y_{M}^{(1)}\, &\D z  +(\frac12-\Delta_\rho) \int_0^{\overline{\zeta}} Y_1 \, \D z\\ 
&=\Drho\big((\rho_1-\rho_3)\alpha_{S,x}^{(3)}+(\rho_{2}-\rho_{3})\alpha_{S,x}^{(4)}\big)+(\frac12-\Drho)(\rho_2-\rho_3)(\alpha_{S,x}^{(3)}+\alpha_{S,x}^{(4)})\\
&=\big(\Drho(\rho_1-\rho_3)+(\frac12-\Drho)(\rho_2-\rho_3)\big)\alpha_{S,x}^{(3)}+\frac12(\rho_2-\rho_3)\alpha_{S,x}^{(4)}\\ & =\frac12(\rho_2-\rho_3)\alpha_{S,x}^{(4)}
\\ 
\int_0^h Y_{M}^{(2)}\, \D z&=(\rho_1-\rho_2)\alpha_{S,x}^{(1)}+\big(\frac{\rho_2-\rho_3}{2}+\Drho\, (\rho_1-\rho_2)\big)\alpha_{S,x}^{(2)}=(\rho_1-\rho_3)\alpha_{S,x}^{(1)}\\
\int_0^\oet Y_{M}^{(2)}\, \D z&=(\rho_2-\rho_3)\alpha_{S,x}^{(1)}+\frac{\rho_2-\rho_3}{2} \alpha_{S,x}^{(2)}\, ,
\end{split}
\end{equation}
where we used definition~\rref{Drrhodef} of $\Drho$ and the fact that the terms with the $z$-derivatives of the Dirac's $\delta$ give a vanishing contribution since they are integrated against functions of $x$ only.

From~(\ref{pushfwd}) we obtain the expression of the reduced Poisson tensor $\Pp$ on $\Ss$, in the coordinates $(\xi_1,\xi_2, \tau_1,\tau_2)$, as
\begin{equation}
\label{Pred}
\Pp=\left(
\begin{array}{cccc} 
0&0&\rho_1-\rho_3&\rho_2-\rho_3\\
0&0&0&\frac12(\rho_2-\rho_3)\\
\rho_1-\rho_3&0&0&0\\
\rho_2-\rho_3&\frac12(\rho_2-\rho_3)&0&0\end{array}
\right)
\partial_x\, .
\end{equation}
The variables $(\xi_1,\xi_2, \tau_1,\tau_2)$ are related to $(\zeta_1,\zeta_2,\sigma_1=\rho_2\ou_2-\rho_1\ou_1,\sigma_2=\rho_3\ou_3-\rho_2\ou_2)$ by
\begin{equation}\label{etatoxi}\begin{split}
\xi_1&=\left( h-\zeta_{{2}} \right)  \left( \rho_{{2}}-\rho_{{3}} \right) +
 \left( h-\zeta_{{1}} \right)  \left( \rho_{{1}}-\rho_{{2}} \right)\,,\qquad \xi_2=\frac12\, \left( \rho_{{2}}-\rho_{{3}} \right)  \left( \zeta_{{1}}-\zeta_{{2
}} \right)\,, \\
\tau_1&=\sigma_1+\sigma_2\,,\qquad
\tau_2=\sigma_2\,.\end{split}
\end{equation}
Solving these relations with respect to the $\zeta_i$'s and the $\sigma_i$'s gives:
\begin{equation}\label{xitoeta}
\begin{split}\medskip
\zeta_1&=-{\frac {\xi_{{1}}}{\rho_{{1}}-\rho_{{3}}}}+{\frac {2\xi_{{2}}}{\rho
_{{1}}-\rho_{{3}}}}+h\,,\qquad
\zeta_2=-{\frac {\xi_{{1}}}{\rho_{{1}}-\rho_{{3}}}}-{\frac {2 \left( \rho_{{
1}}-\rho_{{2}} \right) \xi_{{2}}}{ \left( \rho_{{2}}-\rho_{{3}}
 \right)  \left( \rho_{{1}}-\rho_{{3}} \right) }}+h\,,\\
\sigma_1&=\tau_1-\tau_2\,,\qquad \sigma_2=\tau_2\, .
\end{split}
\end{equation}
A straightforward computation  shows that in these coordinates 
the Poisson operator \rref{Pred} acquires the particularly simple form 
\begin{equation}\label{Preta}
\Pp=\left(
\begin{array}{cccc} 
0&0&-\partial_x&0\\
0&0&0&-\partial_x\\
-\partial_x&0&0&0\\
0&-\partial_x&0&0
\end{array}
\right)\,.
\end{equation}
{\color{black} 
\begin{remark}\label{remarks} 
{
{\em 
According to the terminology favored by the Russian school, for Hamiltonian quasi-linear systems of PDEs  (see, e.g., \cite{Tsa91}) the coordinates $(\xi_1, \xi_2, \tau_1, \tau_2)$ and, {\em a fortiori}, the coordinates $(\zeta_1,\zeta_2, \sigma_1,\sigma_2)$, are ``flat" coordinates for the system. In view of the particularly simple form of the Poisson tensor (\ref{Preta}), the latter set could be called a system of flat {\em Darboux} coordinates. 
}}
\end{remark}
\begin{remark}\label{remarks2}
{\em
 It should be clear how to proceed in the $n$-layered case, with a stratification given by densities $\rho_1<\rho_2<\cdots<\rho_n$ and interfaces
$\zeta_1>\zeta_2>\cdots>\zeta_{n-1}$. We consider intervals
\begin{equation}\label{N-int}
I_1=[0,h],\, I_2=\left[0, \frac{\zeta_1+\zeta_2}{2}\right], I_3=\left[0, \frac{\zeta_2+\zeta_3}{2}\right],\,\dots\,, I_n=\left[0, \frac{\zeta_{n-2}+\zeta_{n-1}}{2}\right]\,,
\end{equation}
and define the analogue of the map (\ref{xitau}) by integrating both $\rho(z)-\rho_n$ and $\sigma(x,z)$ over each of the intervals 
$I_k$, with $k=1,\ldots,n$. The procedure can be repeated {\em verbatim}. We conjecture that, in the long-wave limit, the quantities
\begin{equation}\label{conj-N}
(\zeta_1,\zeta_2,\ldots, \zeta_{n-1}, \sigma_1,\sigma_2,\ldots, \sigma_{n-1}), 
\end{equation}
 where $\sigma_k=\rho_{k+1}\ou_{k+1}-\rho_{k}\ou_{k}$, for $k=1,\ldots, n-1$, are {flat  Darboux coordinates} for the reduced Poisson structure. 
 {We verified that this  conjecture holds true  in the $4$-layer case.}
 \begin{remark}\label{mars}{\em
The steps performed here are borrowed from the Poisson reduction theory of Marsden, Ratiu and Weinstein (see, e.g., \cite{MR86}); however, a more elaborate exposition of our procedure within such a general framework falls beyond the scope of the present work and we omit it here.}
\end{remark}
 }
 \end{remark}
\subsection{The reduced Hamiltonian}
\label{redham}
The full energy (per unit length) of the $2D$ fluid in the channel is just the sum of the kinetic and potential energy, 
\begin{equation}\label{E2D}
H=\int_{-\infty}^{+\infty} \int_0^h \frac12\,\rho 
\left(u^2+w^2\right) \, \D x\, \D z+\int_{-\infty}^{+\infty} \int_0^h  g\rho 
z\, \D x\, \D z\,.
\end{equation}
The potential energy  is readily reduced, using the first of 
\rref{rhouw}, to 
\begin{equation}\label{Ured}
U= \int_{-\infty}^{+\infty}  \frac12 \left(\,g \left( \rho_{{2}}-\rho_{{1}} \right) {\zeta_{{1}}}^{2}+
g \left( \rho_{{3}}-\rho_{{2}} \right) {\zeta_{{2}}}^{2}\right)\, \D x\,.
\end{equation}
When the layer thicknesses are not asymptotically
zero, both energies can be appropriately 
renormalized subtracting the far field contributions of $\zeta_i$. 
To obtain the reduced kinetic energy density, we use the fact that at order~$O(\epsilon^2)$ we can disregard the vertical velocity $w$, and trade the horizontal velocities with their layer-averaged means. Thus the $x$-density is computed as
\begin{equation}\label{Tred}
\mathcal{T}=\frac12\left(\int_0^{\zeta_2} \rho_3 \ou_3^2 \, \D z+\int_{\zeta_2}^{\zeta_1}\rho_2 \ou_2^2\, \D z+\int_{\zeta_1}^h  \rho_1 \ou_1^2\, \D z\right)=\frac12\left( \rho_3{\zeta_2}\ou_3^2 
+ \rho_2({\zeta_1-\zeta_2})\ou_2^2+ \rho_1{(h-\zeta_1)}\ou_1^2\right)\, ,
\end{equation}
so that the reduced kinetic energy is 
\begin{equation}
\label{Tred-bis}
T=\int_{-\infty}^{+\infty} \frac12 \left(\rho_3{\zeta_2}\ou_3^2 
+ \rho_2({\zeta_1-\zeta_2})\ou_2^2+ \rho_1{(h-\zeta_1)}\ou_1^2\right) \D x\, .
\end{equation}
We now use the dynamical constraint {for localized solutions, whereby velocities vanish at infinty,}
\begin{equation}\label{dynconstr}
(h-\zeta_1)\ou_1+(\zeta_1-\zeta_2)\ou_2+\zeta_2\ou_3=0\, , 
\end{equation}
to obtain 
\begin{equation}\label{u1}
\ou_{{1}}={\frac {(\zeta_{{1}}-\zeta_{{2}})\ou_{{2}}+\zeta_{{2
}}\ou_{{3}}}{\zeta_{{1}}-h}}\, .
\end{equation}
Next, we express $\ou_2,\ou_3$ in terms of $\sigma_1,\sigma_2$ as
\begin{equation}\label{u(sigma)}
\begin{split}
\ou_2&=\frac{\rho_{{3}} \left( h-\zeta_{{1}} \right) \sigma_{{1}}}{\Psi}-\frac{\zeta_{{2}}\rho_{{1
}}\sigma_{{2}}}{\Psi}\\
\ou_3&=\frac{\rho_2(h-\zeta_1)\sigma_1}{\Psi}+\frac{\left(h\rho_{{2}}+( \rho_{{1}}-\rho_{{2}}) \zeta_{{1}}-\zeta_{{2}
}\rho_{{1}}\right)\sigma_2}{\Psi}\, , 
\end{split}
\end{equation}
where 
\begin{equation}\label{Psi}
\Psi=h\rho_{{2}}\rho_{{3}}-\rho_{{3}} \left( \rho_{{2}}-\rho_{{1}} \right) \zeta_{{1}}-\rho_{{1}}
 \left( \rho_{{3}}-\rho_{{2}} \right) \zeta_{{2}}\,.
\end{equation}
The kinetic energy density turns out to be, in the new set of variables, 
\begin{equation}\label{Tsigma}
\begin{split}
\mathcal{T}=&{1\over 2\Psi}
\Big(
(h-\zeta_1)\left(\rho_3\zeta_1+ (\rho_2-\rho_3) \zeta_2\right)\sigma_1^2
\\ 
&
+2(h-\zeta_1)\rho_2\,\zeta_2\,\sigma_1\sigma_2+
\left(
(\rho_1-\rho_2) \zeta_2\,\zeta_1+\rho_2\,
\zeta_2 \,h-\rho_1\,\zeta_2^2
\right)
\sigma_2^2 
\Big)\,,
\end{split}
\end{equation}
so that the Hamiltonian functional is 
\begin{equation}\label{FullH}
H=\int_{-\infty}^{+\infty}  \left(\mathcal{T}+\frac12\,g \left(( \rho_{{2}}-\rho_{{1}}) \zeta_1^{2}+
 ( \rho_{{3}}-\rho_{{2}}) \zeta_2^{2}\right) \right)\D x\,.
 \end{equation}
 Explicitly, the equations of motion can be written as conservation laws, 
 \begin{equation}\label{equofmot}
\left( \begin{array}{c}
{\zeta_1}_t\\
{\zeta_2}_t\\
{\sigma_1}_t\\
{\sigma_2}_t \end{array}\right)
=-
\left(\begin{array}{c}
\left(\delta_{\sigma_{1}} H\right)_x\\
\left(\delta_{\sigma_{2}} H\right)_x\\
\left(\delta_{\zeta_{1}} H\right)_x\\
\left(\delta_{\zeta_{2} }H\right)_x
\end{array} \right)\, ,
\end{equation}
where the gradient of the Hamiltonian is, explicitly, 
\begin{equation}\label{gradH}
\begin{split}
& \delta_{\zeta_{1}} H=    
\frac1{2\Psi}\left(\left( h\rho_{{3}}-2\,\zeta_{{1}}\rho_{{3}}+\zeta_{{2}}(\rho_{{3}}-\rho_{{2}})\right) 
{\sigma_{{1}}}^{2}-2\,\zeta_{{2}}\rho_{{2}}
\sigma_{{2}}\sigma_{{1}}-\zeta_{{2}} \left( \rho_{{2}}-\rho_{{1}}
 \right) {\sigma_{{2}}}^{2}\right)\\ &\qquad \qquad -\frac{\rho_3(\rho_1-\rho_2)\, \mathcal{T}}{\Psi}+g \left( \rho_{{2}}-\rho_{{1}} \right) \zeta_{{1}}\,,\\
 &\delta_{\zeta_{2} }H=
 \frac1{2\Psi}
 \left(\left( h-\zeta_{{1}} \right) \left( \rho_{{2}}-\rho_{{3}} \right)  
 {\sigma_{{1}}}^{2}+2\, \left( h-\zeta_{{1}} \right) \rho_{{2}}\sigma_{{2
}}\sigma_{{1}}+ \left( \rho_{{2}}h+\zeta_{{1}}\rho_{{1}}-\zeta_{{1}}\rho_{{2}}-2\,\zeta_{{2}}\rho_{{1}} \right) {\sigma_{{2}}}^{2}
 \right)\\ &\qquad \qquad -\frac{\rho_{{1}} \left( \rho_{{2}}-\rho_{{3}}  \mathcal{T}\right)}{\Psi}+g \left( \rho_{{3}}-\rho_{{2}} \right) \zeta_{{2}}\,,\\
&\delta_{\sigma_{1}} H=\frac1{\Psi} \left(\left( \zeta_{{1}}\rho_{{3}}+\zeta_{{2}}\rho_{{2}}-\zeta_{{2}}\rho_{{3}}
 \right)  \left( h-\zeta_{{1}} \right) \sigma_{{1}}+\rho_{{2}}\zeta_{{2}
} \left( h-\zeta_{{1}} \right) \sigma_{{2}} \right)\,,\\
&\delta_{\sigma_{2}} H= \frac1{\Psi} \left(\rho_{{2}}\zeta_{{2}} \left( h-\zeta_{{1}} \right) \sigma_{{1}}+\zeta_{{2
}} \left( h\rho_{{2}}+\zeta_{{1}}(\rho_{{1}}-\rho_{{2}})-\zeta_{
{2}}\rho_{{1}} \right) \sigma_{{2}}\right) \,.
\end{split}
\end{equation}
As can be seen from the above formulae, even with the simple, constant  Hamiltonian operator~(\ref{Preta}) in~(\ref{equofmot}), this Hamiltonian gradient  leads to  rather lengthy (albeit explicit) expressions for the evolution equations, which are not particularly illuminating and hence are omitted here. {Suffices to say that the conciseness of the Hamiltonian formalism allows to show quickly the existence of at least six conserved quantities,
\begin{equation}
\begin{split} 
&Z_j=\int_{-\infty}^{+\infty}  \zeta_j\, \D x,\qquad S_j=\int_{-\infty}^{+\infty}  \sigma_j\, \D x, \quad j=1,2\\
&H\> \text{ given by equation (\ref{FullH})},\quad K=\int_{-\infty}^{+\infty}  (\zeta_1\sigma_1+\zeta_2\sigma_2) \, \D x\, .\end{split}
\end{equation}
In particular, the quantity $K$ indicates how  
the momentum paradox mentioned in Section~\ref{Momparad} can be viewed from the canonical formulation of the evolution equations.  In fact, we 
see that, rather than the horizontal momentum $\Pi^{(x)}$,  it is $K$ that plays the role of generator of $x$-translations.
By expressing its density $\mathcal{K}=\zeta_1\sigma_1+\zeta_2\sigma_2 $ in terms of the horizontal mean velocities we obtain
\begin{equation}\label{xgenu}
\mathcal{K}=\rho_{{2}}\ou_{{2}} \left( \zeta_{{1}}-\zeta_
{{2}} \right) +\rho_{{3}}\ou_{{3}}\zeta_{{2}}-\rho_{{1}}\ou_{{1}}\zeta_{{1}}=\pi^{(x)}-h\rho_1\ou_1\,,
\end{equation}
where
\[
\pi^{(x)}=\left( h-\zeta_{{1}} \right) \rho_{{1}}\ou_{{1}}+\rho_{{2}}\ou_{{2}}
 \left( \zeta_{{1}}-\zeta_{{2}} \right) +\rho_{{3}}\zeta_{{2}}\ou_{{3}} \, 
\]
is the total horizontal momentum density. Thus, the total momentum 
$\Pi^{(x)}=\int_{-\infty}^{+\infty}   \pi^{(x)}\, \D x$
is {\em not} conserved, while $K=\int_{-\infty}^{+\infty}  \mathcal{K} \D x$ is.}
\section{The Boussinesq approximation}
A dramatic simplification of the problem is provided by the so-called {\em Boussinesq approximation}, that is, 
the double scaling limit 
 \begin{equation}\label{dslim-bis}
\rho_i\to  \orho=\frac{\rho_1+\rho_2+\rho_3}{3},\quad i=1,2,3, \quad \text{with $g(\rho_1-\rho_2)$ and $g(\rho_2-\rho_3)$ both finite.} 
\end{equation}
Since the Poisson tensor~(\ref{Preta}) is independent of the densities, this double scaling limit can be implemented most simply within the Hamiltonian formulation. While the potential energy is unchanged, from \rref{Tsigma} the kinetic energy acquires  the form
\begin{equation}\label{KinB}
{\frac { \left( h-{\zeta_{{1}}}\right)\zeta_{{1}} {\sigma_{{1}}}^{2}}{2h{\orho}}}+{\frac {\zeta_{{2}} \left( h-\zeta_{{1}} \right) \sigma_{{2}}\sigma_{{1}
}}{h\orho}}+{\frac {\zeta_{{2}} \left( h-\zeta_{{2}} \right) {\sigma_{{2}}}^{2}}{2h{\orho}}}\,,
\end{equation}
 so that the Hamiltonian energy  functional in this Boussinesq limit is 
\begin{equation}\label{HamB}
\begin{split}
H_B=&\int_{-\infty}^{+\infty} 
\frac1{2\, h\orho} \left(
{ { \zeta_1(h-\zeta_{{1}}){\sigma_{{1}}}^{2}}}+
2{ {\zeta_{{2}} \left( h-\zeta_{{1}} \right) \sigma_{{1}}\sigma_{{2}}}}
+{ {\zeta_{{2}} \left( h-\zeta_{{2}} \right) {{\sigma_2}^2}}}
\right)\, 
\\ &
+g(\left( \rho_{{2}}-\rho_{{1}} \right) {\zeta_{{1}}}^{2}+
 \left( \rho_{{3}}-\rho_{{2}} \right) {\zeta_{{2}}}^{2})
 \,\D x\,. \end{split}
\end{equation}
The ensuing equations of motion are 
\begin{equation}
\label{HaeqB}
\left( \begin{array}{c}
{\zeta_1}_t\\
{\zeta_2}_t\\
{\sigma_1}_t\\
{\sigma_2}_t \end{array}\right)
=\mathcal{P} 
\left(\begin{array}{c}
\delta_{\zeta_{1}} H_B\\
\delta_{\zeta_{2}} H_B\\
\delta_{\sigma_{1}}H_B\\
\delta_{\sigma_{2} }H_B
\end{array} \right)\,,
\end{equation}
with $\mathcal{P}$ the ``canonical" Poisson tensor (\ref{Preta}).
As a system of quasi-linear equations, they can be cast in the form
\begin{equation}\label{Boussieqham-bis}
\left( \begin{array}{c}
{\zeta_1}_t\\
{\zeta_2}_t\\
{\sigma_1}_t\\
{\sigma_2}_t \end{array}\right)+
\mb{A}
\left( \begin{array}{c}
{\zeta_1}_x\\
{\zeta_2}_x\\
{\sigma_1}_x\\
{\sigma_2}_x \end{array}\right)=0\, ,
\end{equation}
where the charateristic matrix reads
\begin{equation}\label{BouMat-bis}
\mb{A}=
\frac1{h\, \orho}\left(\begin{array}{cccc}\medskip
\begin{array}{l} { \left( 2\,\zeta_1 -h\right) \sigma_1}\\ +{\sigma_2\zeta_2}\end{array} &
{ \left( \zeta_{{1}} -h \right) \sigma_{{2}} }&\zeta_{{1}} \left( \zeta_{{1}} -h \right)& \zeta_{{2}} \left( \zeta_{{1}} -h \right)\\ \medskip
\zeta_{{2}}\sigma_1& \begin{array}{l}\left( \zeta_{{1}} -h\right) \sigma_{{1}}\\+\left( 2\,\zeta_{{2}}-h \right) \sigma_{{2}}\end{array}&
\zeta_{{2}} \left( \zeta_{{1}}-h \right)&\zeta_{{2}} \left( \zeta_{{2}} -h\right)\\ \medskip
\sigma_{{1}}^{2}+\tilde{g} \left( \rho_{{1}}-\rho_{{2}} \right) &
\sigma_{{1}}\sigma_{{2}}&\begin{array}{l}\left( 2\,\zeta_{{1}}-h \right) \sigma_{{1}}\\+\sigma_{{2}}\zeta_{{2}}\end{array}&\zeta_{{2}}\sigma_{{1}}\\ \medskip
\sigma_{{1}}\sigma_{{2}}&\sigma_{{2}}^{2}+\tilde{g} \left( \rho_{{2}}-\rho_{{3}} \right) &\left( \zeta_{{1}}-h \right) \sigma_{{2}}
&\begin{array}{l} \left( \zeta_{{1}}-h \right) \sigma_{{1}}\\+\left(2\,\zeta_{{2}} -h\right) \sigma_{{2}}\end{array}
\end{array}\right)
\end{equation}
and $\tilde{g} =g\, h\, \orho$ is the reduced gravity.

As in the free surface case,  the question of existence of Riemann invariants for this quasi-linear system can be easily answered by computing (by means of standard computer algebra programs) the Haantjes tensor $\mathcal{H}$ of the matrix $\mb{A}$.
A lengthy computation shows that $\mathcal{H}$ has $12$ (out of $24$) non-vanishing components, namely
\begin{equation}
\label{Haantcomp}\begin{split}
&{\HH^1}_{1\, 2}={\HH^3}_{2\, 3}=-{\HH^4}_{1\, 3}=-{\frac {g\,r_{{32}}\,\zeta_{{2}} \left( h-\zeta_{{1}} \right) {\sigma_{{1}}
}^{2}}{{\bar\rho}^{3}{h}^{2}}}+{\frac {g\,r_{{21}}\, \left( \zeta_{{1}}-2\,\zeta
_{{2}} \right)  \left( h-\zeta_{{1}} \right) {\sigma_{{2}}}^{2}}{{\bar\rho}
^{3}{h}^{2}}}+{\frac {{g}^{2}r_{{21}}r_{{32}} \zeta_{{2}} \left( h-\zeta_
{{1}} \right) }{{\bar\rho}^{2}h}}
\\ 
& {\HH^2}_{1\, 2}={\HH^3}_{2\, 4}=-{\HH^4}_{1\, 4}=-{\frac {g\, r_{{32}}\,\zeta_{{2}} \left( \zeta_{{2}}+h-2\,\zeta_{{1}}
 \right) {\sigma_{{1}}}^{2}}{{\bar\rho}^{3}{h}^{2}}}-{\frac {g\, r_{{21}}\, \zeta_{{2}} \left( h-\zeta_{{1}} \right) {\sigma_{{2}}}^{2}}{{\bar\rho}^{3}{h}^{
2}}}+{\frac {{g}^{2}\, r_{{21}}r_{{32}}\, \zeta_{{2}} \left( h-\zeta_{{1}}
 \right) }{{\bar\rho}^{2}h}}\\ 
 &{\HH^1}_{1\, 4}=-{\HH^2}_{1\, 3}=-{\HH^3}_{3\, 4}=2\,{\frac {g\, r_{{21}}\zeta_{{2}}\sigma_{{2}} \left( \zeta_{{1}}-\zeta_{{2}
} \right)  \left( h-\zeta_{{1}} \right) }{{\bar\rho}^{3}{h}^{2}}}
\\
& {\HH^1}_{2\, 4}=-{\HH^2}_{2\, 3}=-{\HH^4}_{3\, 4}=
-2\,{\frac {g\, r_{{32}}\,\zeta_{{2}}\sigma_{{1}} \left( \zeta_{{1}}-\zeta_{{2
}} \right)  \left( h-\zeta_{{1}} \right) }{{\bar\rho}^{3}{h}^{2}}}\, 
\end{split}
\end{equation}
where $r_{ij}=\rho_i-\rho_j$.
As should be expected from the free surface case, even in the Boussinesq approximation the model for $3$-layer fluid confined between rigid surfaces does not admit Riemann invariants. As can also be expected, this non-existence extends~{\em a fortiori\/} to the general non-Boussinesq system, as well as to~$n$-layered models with a rigid lid  for $n > 3$. {The implications on the structural properties of quasi-linear systems that do not admit Riemann invariants briefly discussed in Remark~\ref{Remaref2}  naturally apply to all these cases as well.}

\subsection{Symmetric solutions}
In the recent paper \cite{VirMil19} the authors have focussed on the symmetric solutions defined by the equality of the upper ($i=1$) and lower ($i=3$) 
layer thicknesses, i.e., $\zeta_2=h-\zeta_1$, and the averaged horizontal velocities, $\ou_1=\ou_3$. In the Boussinesq approximation, our variables 
$(\sigma_1,\sigma_2)$ are actually proportional to the velocity shears, 
\begin{equation}\label{sigmau}
\sigma_1=\orho (\ou_2-\ou_1),\quad \sigma_2=\orho (\ou_3-\ou_2)\, , 
\end{equation}
 so that  the symmetric solutions found in~\cite{VirMil19} are given by the relations
 \begin{equation}\label{symm}
\zeta_2=h-\zeta_1,\quad \sigma_2=-\sigma_1\,.
\end{equation}
A straightforward computation confirms that the submanifold defined by these relations is invariant under the flow \rref{Boussieqham-bis} if and only if the relation
\begin{equation}\label{symeq}
\rho_3-\rho_2=\rho_2-\rho_1
\end{equation}
is fulfilled among the density differences.
In this case system (\ref{Boussieqham-bis}) reduces to a system with $2$~``degrees of freedom," parametrized, e.g., by the pair $(\zeta_2\equiv\zeta, \sigma_2\equiv \sigma)$. The reduced ``symmetric" equations of the motion are \begin{equation}\label{redeq2}
\left(
\begin{array}{c}
\zeta_{t}\\
\sigma_t\end{array}
\right)=\frac1{\orho h}
\left(\begin{array}{cc}
\left(4\,\zeta- \,h\right)\sigma & { { \zeta   
\left( 2\,\zeta -h\right) }}\\ 
\noalign{\medskip}
2\,{\sigma}^{2}-g\orho\drho h & \left(4\,\zeta- \,h\right)\sigma
\end{array}\right)
\left(
\begin{array}{c}
\zeta_{x}\\
\sigma_x\end{array}
\right)\, , 
\end{equation}
where we have defined $\drho=\rho_3-\rho_2$.
These equations follow from the Hamiltonian functional
\begin{equation}\label{H2}
H_{B,S}=\int_{-\infty}^{+\infty}
\left({\frac {  \zeta   \left(h- 2\,\zeta \right) 
{\sigma }^{2}}{2\orho h}}+\frac12\,g\drho \left(\zeta-{h\over 2}  \right)^2 \right) \D x\, ,
\end{equation}
with the ``standard" Poisson tensor
\begin{equation}\label{P2op}
\mathcal{P}_{(2)}=\left(\begin{array}{cc} 0&-\partial_x\\ -\partial_x&0\end{array}\right)\, , 
\end{equation}
where the reference level for the potential energy in the Hamiltonian density is chosen at the midpoint of the channel. 
From equations \rref{redeq2} the characteristic velocities of the system can be read off immediately, 
\begin{equation}\label{charspeed}
\lambda_\pm=\frac{1}{\orho h} \left( \sigma(h-4\zeta)\pm\sqrt{  \zeta  \left(h-2\,\zeta \right)  \left( 
g h\orho \drho-2\,{\sigma }^{2} \right)}\right)\, ; 
\end{equation}
since in this case $\zeta\in(0,{h}/2)$, the domain of hyperbolicity  coincides with the rectangular region 
\[
\left(0,\frac{h}2\right)\times \left(\dsl{-\frac1{\sqrt{2}}\sqrt{gh\orho\drho}},  \dsl{\frac1{\sqrt{2}}\sqrt{gh\orho\drho}}\right)\, .
\]
Alternatively, the Hamiltonian formulation shows that the simple map $\zeta \to \eta$, $\sigma \to h \sigma$, $\orho\to \rho_2$, $\drho \to\rho_2-\rho_1$ turns the Hamiltonian in~(\ref{H2}) and the operator in~(\ref{P2op}) into that of the two-layer Boussinesq system in a channel of half width $h/2$ reported in~\cite{CFOPT19} (as also remarked in~\cite{VirMil19}, though only reported explicitly from the motion equation viewpoint in the Ph.D. thesis~\cite{VirThes}), thereby translating all the properties discussed therein to the present three-layer Boussinesq symmetric case. In fact, 
{\color{black}
these symmetric solutions have a clear meaning in the Hamiltonian setting.
Consider the involution $\mathcal{J}$ defined by
\begin{equation}\label{invo}
\widetilde{\zeta}_1=h-\zeta_2\,,\quad 
\widetilde{\zeta}_2=h-\zeta_1\,,\quad 
\widetilde{\sigma}_1=-\sigma_2\,,\quad 
\widetilde{\sigma}_2=-\sigma_1\,.
\end{equation}
Since its Jacobian is
\begin{equation}
\mathcal{J}_*=\left(
\begin{array}{cccc} 0&-1&0&0\\
-1&0&0&0\\
0&0&0&-1\\
0&0&-1&0
\end{array}
\right)\, ,
\end{equation}
the Poisson tensor (\ref{Preta}) is preserved by the involution. As far as the Hamiltonian (\ref{HamB}) is concerned, the kinetic energy density is invariant under the involution (\ref{invo}), while the potential energy density, when $\rho_3-\rho_2=\rho_2-\rho_1\equiv \rho_\Delta$, changes by a constant term added to the linear term $H_\Delta=-g\rho_\Delta\, (\widetilde{\zeta}_1+\widetilde{\zeta}_2)$. Since 
$
\int_{-\infty}^{+\infty} H_\Delta\, \D x
$ 
is a Casimir function for the Poisson tensor (\ref{Preta}), the equations of motion are left invariant by the involution (\ref{invo}). In other words, symmetric solutions correspond to the manifold of fixed points of the {\em canonical} involution (\ref{invo}), and  $H_{B,S}$ is just the restriction of the Hamiltonian $H_B$ (up to a factor $1/2$) to the space defined by relations~(\ref{symm}) under conditions~(\ref{symeq}) for the density differences.

%
%

 \section{Summary and conclusions}
 
 In this paper we examined some structural properties of multi-layered incompressible Euler flows in the long-wave regime.  In order to present explicit computations, we mainly focussed on the three-layer case. The aim was to  point out similarities and differences with the two-layer setups which has received much attention in the literature
 (see, e.g., \cite{Chumaetal09,CCFOP12}).
 
 At first, we briefly recalled how effective equations in one space dimension are obtained by layer-means and the hydrostatic approximation for the pressure in the dispersionless case. We then recalled some known facts about the free surface case, highlighting the natural Hamiltonian structure of the ensuing equations, and the non-existence of the Riemann invariants for the quasi-linear resulting system, a fact proven in \cite{Eletal17}.
 
Next, we moved onto the main focus of the present paper, the case of stratified fluid in a vertical channel, a configuration that mathematically translates to the enforcement of the rigid lid upper constraint. 
 The phenomenon of effective pressure differentials implying the ``paradox" of non-conservation of the horizontal momentum, highlighted in \cite{CCFOP12} for the two-layer case, has been shown to persist for an $n$-layered situation, $n\ge 3$, and in fact even be enhanced, for zero initial velocities,  by scaling linearly with density differences, as opposed to quadratically as in the two layer case. A natural Hamiltonian  structure on the configuration space of effective (in one space dimension)  $3$-layered fluid motions was derived by means of a geometrical reduction process from the full $2$-dimensional Hamiltonian structure introduced in \cite{Ben86}. This allowed us to write the equations of motion in the form of a system of conservation laws, and to recover the correct conserved quantity (the impulse) associated with the translational Noether symmetry of the system. 
 
The Hamiltonian formulation led to a simple derivation of the Boussinesq limit of the motion equations. The (expected) non-existence of Riemann invariants in the rigid lid case was proved, and the geometrical meaning of the ``symmetric" configurations recently studied in \cite{VirMil19} is pointed out. The determination of invariant submanifolds of the full equations ensuing from the Hamiltonian~(\ref{FullH}), conditioned by suitable constraints on the density differences, and the study of the properties of such reduced systems is a non trivial question. Its analysis is left for future investigations.}

Future developments will include the extension of our formalism to the next order in the long-wave parameter asymptotics, with the inclusion in this framework of the dispersive terms $D_i$ of the $n$-layer 
equations (\ref{nlayer}). Framing this problem within Dubrovin's approach of the expansion of Hamiltonian PDEs in the dispersion parameter (see, e.g., \cite{Du08, DGKM14}) is a task worth pursuing, as well as the comparison of our setting with that of~\cite{PCH}.

\par\smallskip\noindent
{\bf Acknowledgments.}
This project has received funding from the European Union's Horizon 2020 research and innovation programme under the Marie Sk{\l}odowska-Curie grant no 778010 {\em IPaDEGAN}. We also gratefully acknowledge the auspices of the GNFM Section of INdAM under which part of this work was carried out. RC thanks the support by the National Science Foundation under grants RTG DMS-0943851, CMG ARC-1025523, DMS-1009750, DMS-1517879, and by the Office of Naval Research under grants N00014-18-1-2490 and  DURIP N00014-12-1-0749. MP thanks  the {Dipartimento di Matematica e Applicazioni} of Universit\`a Milano-Bicocca  for its hospitality. All authors gratefully thank the anonymous referees for their comments and remarks. 

{
\section*{Appendix: Hamiltonian structure in the free surface 3-layer case}
\renewcommand{\theequation}{A\arabic{equation}}
\label{AppA}
The effective equations of motion in the general case of $n$-layer stratified fluids with a free upper surface is given by system~(\ref{mean-hydro-Euler}) with the pressure at the bottom being $P^{0}(x)=g\sum_{k=1}^{n}\rho_{k}\eta_{k}$. 
These possess a natural Hamiltonian structure, which we illustrate here  with the special case of a 3-layered fluid. 
We denote by $\eta_{i}$ the thickness of the $i$-th layer, by $z=\zeta_{0}$ the free surface, by $z=\zeta_{1}$ and $z=\zeta_{2}$  the upper interface and  lower interface, respectively,  and by $z=\zeta_{3}=0$ the bottom of the channel, as illustrated by Figure~\ref{freesurf3}.
\begin{figure}[t]
\begin{center}\includegraphics[width=13cm]{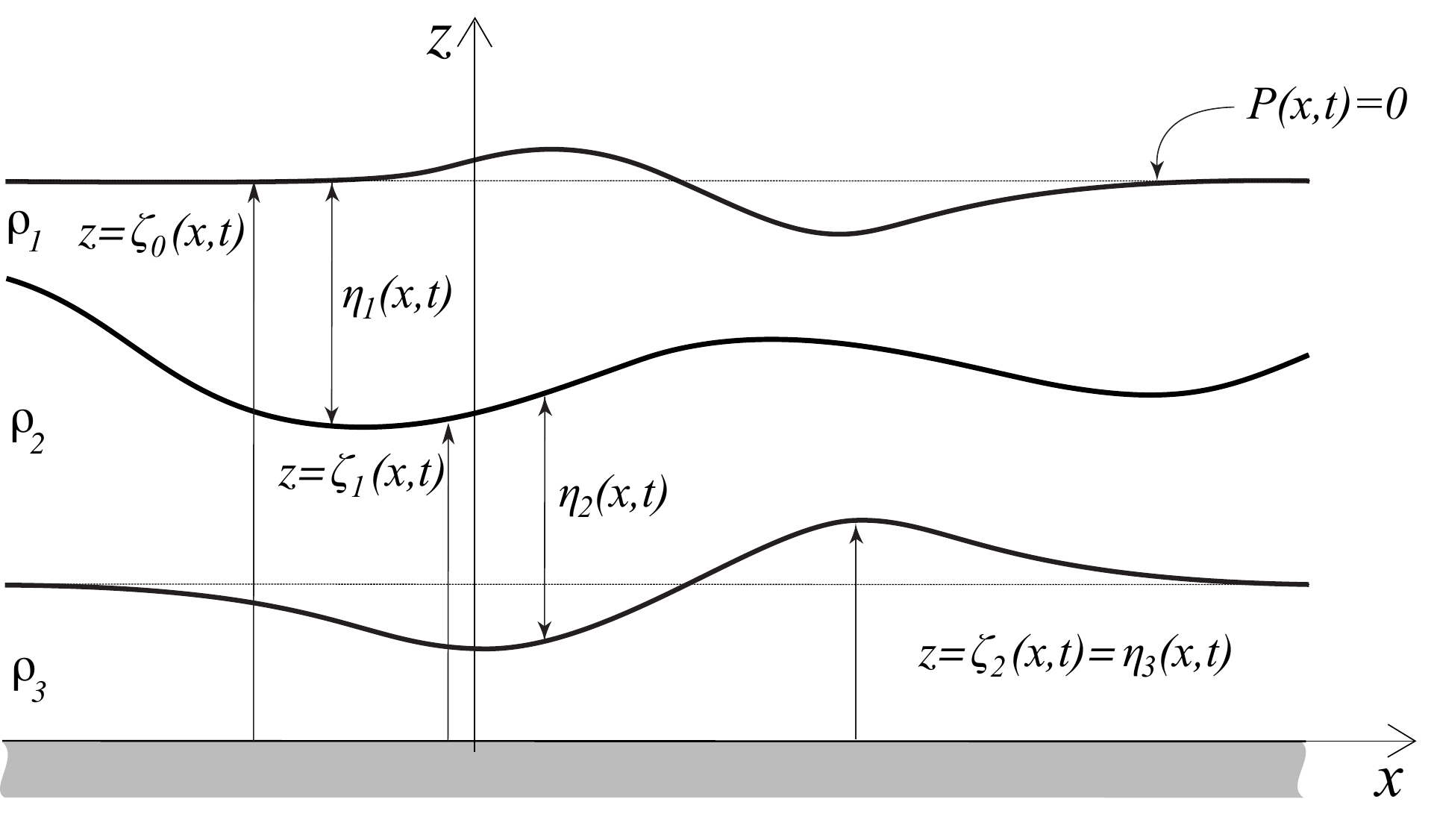}
\end{center}
\caption{Schematic of the free surface, three-layer stratified fluid set up; notation adapted from Figure~\ref{Z}.}
\label{freesurf3}
\end{figure}
The interfacial pressures are, respectively, 
$$ p_{3}(x,z)=P^{0}(x)-\rho_{3}gz\,,$$
for the lowest layer $0<z<\zeta_{2}$, 
$$ p_{2}(x,z)=P^{0}(x)-g\rho_{3}\eta_{3} -\rho_{2}g(z-\zeta_{2}) \,,  $$
for the middle layer $\zeta_2<z<\zeta_{1}$,
$$ p_{1}(x,z)=P^{0}(x)-g(\rho_{3}\eta_{3}+\rho_{2}\eta_{2})-\rho_{1}g(z-\zeta_{1})\,, $$
for the first layer $\zeta_1<z<\zeta_{0}$.
The corresponding layer-mean pressures are $\overline{p}_{i,x}=p_{ix}$ for $i=1,2,3$, that is, 
\begin{align*}
p_{1x}&=g\rho_{1}(\eta_{1x}+\eta_{2x}+\eta_{3x}) \,,\\
p_{2x}&=g(\rho_{1}\eta_{1x}+\rho_{2}\eta_{2x}+\rho_{2}\eta_{3x})\, ,\\
p_{3x}&=P^{0}_{x}=g(\rho_{3}\eta_{3x}+\rho_{2}\eta_{2x}+\rho_{1}\eta_{1x})\,.
\end{align*}
Plugging these expressions in the general formula (2.10), we obtain the equations of motion for the 3-layered stratified fluid with free upper surface
\begin{align*}
(\eta_{it})+(\eta_{i}\overline{u}_{i})_{x}=0\, \qquad  \text{ for } i=1,2,3\\\nonumber
\overline{u}_{1t}+\overline{u}_{1} \overline{u}_{1x}+g(\eta_{1x}+\eta_{2x}+\eta_{3x})=0 \,,\\
\overline{u}_{2t}+\overline{u}_{2}\overline{u}_{2x}+g\left(\dfrac{\rho_{1}}{\rho_{2}} \eta_{1x}+\eta_{2x}+\eta_{3x}\right)=0\,,\\
\overline{u}_{3t}+\overline{u}_{3}\overline{u}_{3x}+g\left(\eta_{3x}+\dfrac{\rho_{2}}{\rho_{3}}\eta_{2x}+\dfrac{\rho_{1}}{\rho_{3}}\eta_{1x} \right)=0 \,.
\end{align*}
To show that these equations admit a natural Hamiltonian formulation, the averaged velocities $\overline{u}_{i}$ can be replaced by  the averaged momentum coordinates $\overline{\mu}_{i}$ defined by 
$\overline{\mu}_{i}=\rho_{i}\overline{u}_{i}$,  $i = 1, 2, 3$, so that 
\begin{align}
\eta_{it}+\dfrac{1}{\rho_{i}}(\eta_{i}\overline{\mu}_{i})_{x}=0\,, \text{ for } i=1,2,3, \\ \nonumber
\overline{\mu}_{1t}+\dfrac{1}{\rho_{1}}\overline{\mu}_{1} \overline{\mu}_{1x}+g\rho_{1}(\eta_{1x}+\eta_{2x}+\eta_{3x}) =0 \,, \nonumber\\
\overline{\mu}_{2t}+\dfrac{1}{\rho_{2}}\overline{\mu}_{2}\overline{\mu}_{2x}+ g(\rho_{1}\eta_{1x}+\rho_{2}\eta_{2x}+\rho_{2}\eta_{3x})=0\,, \\
\overline{\mu}_{3t}+\dfrac{1}{\rho_{3}}\overline{\mu}_{3}\overline{\mu}_{3x}+ g(\rho_{3}\eta_{3x}+\rho_{2}\eta_{2x}+\rho_{1}\eta_{1x})=0 \,. \nonumber \\
\nonumber
\end{align}
It is straightforward to check that these equations can be written in Hamiltonian  form 
\begin{equation}
\left( \begin{array}{c}
{\eta _{1t}}\\
{\eta _{2t}}\\
{\eta _{3t}}\\
{\overline{\mu} _{1t}}\\
{\overline{\mu}_{2t}}\\
{\overline{\mu}_{3t}}
\end{array} \right) = \left( \begin{array}{cccccc}
0&0&0&{ - {\partial _x}}&0&0\\
0&0&0&0&{ - {\partial _x}}&0\\
0&0&0&0&0&{ - {\partial _x}}\\
{ - {\partial _x}}&0&0&0&0&0\\
0&{ - {\partial _x}}&0&0&0&0\\
0&0&{ - {\partial _x}}&0&0&0
\end{array} \right)\left( \begin{array}{*{20}{c}}
{{\delta _{{\eta _1}}}H}\\
{{\delta _{{\eta _2}}}H}\\
{{\delta _{{\eta _3}}}H}\\
{{\delta _{{\overline{\mu}_1}}}H}\\
{{\delta _{{\overline{\mu}_2}}}H}\\
{{\delta _{{\overline{\mu}_3}}}H}
\end{array} \right)\,,
\end{equation}
with the natural Hamiltonian functional $H=T+V$ given by
%
%
%
%
%
%
\begin{equation}
H= \int_{-\infty}^{+\infty} \,\frac{1}{2} \left( \dfrac{\eta_{3}}{\rho_{3}}\overline{\mu}_{3}^{2} + \dfrac{\eta_{2}}{\rho_{2}}\overline{\mu}_{2}^{2}   +\dfrac{\eta_{1}}{\rho_{1}}\overline{\mu}_{1}^{2} + g \left(\rho_{3}\eta_{3}^{2}+\rho_{2}\eta_{2}^{2}+\rho_{1}\eta_{1}^{2}+2\rho_{2}\eta_{2}\eta_{3}+2\rho_{1}\eta_{1}\,\eta_{2}+2\rho_1\eta_1\eta_{3}\right)\right) \D x \,.
\end{equation}}


\end{document}